\documentclass[final,3p,times,twocolumn]{elsarticle}

\usepackage{amssymb}
\usepackage{subfig}
\usepackage{amsmath}
\usepackage{algorithmic}
\usepackage{makecell}
\usepackage[ruled,linesnumbered]{algorithm2e}

\journal{Computer Networks}

\newtheorem{definition}{Definition}
\newtheorem{proof}{Proof}
\newtheorem{example}{Example}

\makeatletter
\newcommand{\removelatexerror}{\let\@latex@error\@gobble}
\makeatother

\begin{document}

\begin{frontmatter}



\title{PoVF: Empowering Decentralized Blockchain Systems with \\Verifiable Function Consensus} 

\author[authors-uestc]{Chenxi Xiong}
\ead{202221081027@std.uestc.edu.cn}
\author[authors-uestc]{Ting Yang\corref{cor1}}
\ead{yting@uestc.edu.cn}
\author[authors-uestc]{Yu Wang}
\ead{13658087747@163.com}
\author[authors-uestc]{Bing Dong}
\ead{dongbinguestc@gmail.com}
\cortext[cor1]{Email address: yting@uestc.edu.cn}
\affiliation[authors-uestc]{organization={University of Electronic Science and Technology of China},
            city={Chengdu},
            postcode={611731},
            state={Sichuan},
            country={China}}

\begin{abstract}
    Consensus mechanism is the core technology for blockchain to ensure that transactions are executed in sequence. It also determines the decentralization, security, and efficiency of blockchain. Existing mechanisms all have certain centralization issues and fail to ensure the decentralization of blockchain networks. A decentralized and efficient mechanism is required to improve blockchain systems. This paper proposes a fair consensus mechanism called Proof of Verifiable Functions (PoVF), based on the verifiability and unpredictability of verifiable functions. PoVF provides a sufficiently fair mechanism, ensuring that all nodes in blockchain network have equal opportunity to participate in consensus. In addition, a structure called "Delay buffer" is proposed to ensure transactions are executed sequentially. It delay the selection of blocks to avoid blockchain forks caused by broadcasting and transaction execution confusion. According to our security analysis, PoVF is provably secure and has the ability to resist potential adversaries. According to the experiments, PoVF-based blockchain can process up to $4 \times 10^3$ transactions per second with nodes configured with only 4-core CPUs. This paper uses the Gini coefficient to measure the decentralization of blockchains, and the PoVF-based blockchain achieves the lowest Gini coefficient of 0.39 among all sampled blockchains. PoVF has been shown to provide sufficient efficiency while ensuring decentralization and security through experiments.
\end{abstract}




\end{frontmatter}



\section{Introduction}
\label{introduction}
Since the Bitcoin \cite{nakamoto2008bitcoin} proposed by Satoshi Nakamoto, blockchain has attracted considerable attention due to the significant benefits it can offer. The goal of blockchain is to provide decentralized, non-tampering, non-forgery, and traceability storage for important anti-counterfeiting data. This makes blockchain widely applicable in fields such as Internet of Vehicles (IoV) \cite{feng2024pbag} and Internet of Things (IoT) \cite{wang2020blockchain}, etc. The consensus mechanism is the core technology for achieving these benefits in blockchain. It ensures that transactions are executed in sequence and prevents double-spending problem \cite{chohan2021double}. In addition, consensus mechanisms determine the decentralization, security, and efficiency of blockchain. However, existing mechanisms have not been able to achieve sufficient decentralization. Ensuring decentralization and efficient remains a challenge in blockchain \cite{huang2021rich}.

The first consensus mechanism introduced in Bitcoin is known as Proof of Work (PoW) \cite{nakamoto2008bitcoin}. PoW is a consensus mechanism based on hash functions. It requires nodes to solve a computation-intensive hash-based puzzle to achieve consensus. This was the original idea of Bitcoin - \textit{One CPU one vote}. All nodes in the blockchain network have the opportunity to participate in consensus, ensuring the decentralization of Bitcoin. However, with the advent of mining pools, the Bitcoin blockchain started to become centralized \cite{sai2021taxonomy, ren2019pooled}. At the same time, large-scale mining also leads to excessive energy waste, which is the most criticized issue of the Bitcoin blockchain system.

\begin{example}
  In PoW-based blockchain networks, all nodes accept only blocks whose hash values are below the current target threshold. In Bitcoin, the target threshold dynamically adjusts during the mining process to ensure that a valid block is generated approximately every 10 minutes across the network. Miners iteratively modify the nonce in the new block to attempt to produce a hash value that meets the target threshold. Therefore, a consensus mechanism serves as the "rules of the game" in a blockchain, guiding network nodes on how subsequent blocks are generated and selected. This enables the blockchain network to achieve self-management in a manner distinct from Mobile ad-hoc networks (MANETs) \cite{wheeb2020simulated}.
\end{example}

To resolve the centralization and energy waste issues in PoW, Proof of Stake (PoS) was proposed in Peercoin \cite{king2012ppcoin}. The core in PoS is that users holding more coins and longer periods can receive more rewards \cite{shifferaw2021limitations}. However, there is a tendency for coins to aggregate among users with larger holdings, resulting in another form of centralization within blockchain system. Futhermore, PoS is more susceptible to malicious attacks including long-range attack \cite{deirmentzoglou2019survey}, stake-bleeding attack \cite{gavzi2018stake}, etc. Thus, PoS dose not solve the centralization issue, instead it brings more security issues. Both PoS and mechanisms improved based on PoS share similar issues in that they are not sufficiently decentralized. In summary, a sufficiently decentralized and secure consensus mechanism is needed in blockchain technology.

In this paper, we introduce PoVF, a decentralized consensus algorithm based on verifiable functions. PoVF sufficiently leverages the unpredictability of verifiable functions, ensuring that all nodes in the blockchain network have equal opportunity to participate in consensus. According to the security analysis, PoVF is provably secure, and it can efficiency prevent potential adversaries. A PoW-like heartbeat mechanism is introduced in PoVF, which ensures that each node can only have a single identity to effectively prevent Sybil attacks. In addition, PoVF can provide sufficient transaction process speed according to the experiment result. The maximum transactions per second (TPS) in PoVF-based blockchain system\footnote{https://github.com/chain-lab/go-norn} can reach over $4 \times 10^3$ with 200 nodes in blockchain network. PoVF provides a secure and decentralized consensus mechanism that ensures efficient transaction processing within tolerable latency.

The remaining paper is organized as follows. Section \ref{related-work} provides an overview of existing blockchain consensus mechanisms and discusses their limitations with respect to decentralization. Section \ref{preliminaries} introduces the algorithms essential for PoVF. Section \ref{povf-consensus} provides a comprehensive exposition of the PoVF process, representing the central contribution of this paper. Section \ref{security-analysis} presents the security proofs for PoVF, demonstrating its provable security. Section \ref{evaluation} conducts throughput tests on the PoVF-based blockchain, demonstrating that PoVF is sufficiently efficient. Additionally, by measuring the Gini coefficient and standard deviation under different consensus mechanisms, it analyzes and show the superior decentralization of PoVF. Section \ref{conclusion} provides a summary of the paper and outlines directions for future research.

\section{Related Work}
\label{related-work}

Consensus mechanisms in blockchain are used to ensure the consistency of node data. PoW is the first consensus algorithm applied to blockchain technology. The main idea of PoW is solving computation-intensive hash-based puzzles. The issues with PoW and similar proposals are energy waste and low throughput. To improve throughput, Conflux \cite{li2020decentralized} points out another issue in Bitcoin, where the serialization process results in forked blocks, which not only reduces network security but also decrease the efficiency of blockchain system. Thus, a new blockchain structure based on directed acyclic graphs (DAG) is proposed, which effectively utilizes forked blocks to increase throughput. Nevertheless, Conflux still relies on the PoW consensus algorithm, which means it faces the same issues as PoW including energy waste and centralization caused by mining pools.

\begin{table*}[htbp]
  \caption{Consensus mechanisms comparison}
  \begin{center}
  \begin{tabular}{cm{3cm}cccc}
  \hline
  \rule{0pt}{8pt}
  Mechanism & Blockchain & Energy consumption& Matthew effect & Randomness & Centralization \\
  \hline
  \rule{0pt}{8pt}
  PoW & Bitcoin, Conflux& high & low & low & high\\
  \hline
  \rule{0pt}{8pt}
  PoS & Peercoin, Ethereum & low & high & low & high\\
  \hline
  \rule{0pt}{8pt}
  DPoS & Cardano &low & high & low & high\\
  \hline
  \rule{0pt}{8pt}
  PBFT & Hyperledger Fabric &low & low & low & low\\
  \hline
  \rule{0pt}{8pt}
  PoH & Solana & low & high & low & low\\
  \hline
  \rule{0pt}{8pt}
  VDC & - & low & low & high & high\\
  \hline
  \rule{0pt}{8pt}
  PoV & - &low & low & high & high\\
  \hline
  \rule{0pt}{8pt}
  R3V & - &low & low & high & unknown\\
  \hline
  \rule{0pt}{8pt}
  VPoS & Algorand &low & high & medium & high\\
  \hline
  \rule{0pt}{8pt}
  PoVF & - &low & low & high & low\\
  \hline
  \end{tabular}
  \label{related-works-comparison}
  \end{center}
\end{table*}

Proof of Stake (PoS) was proposed in Peercoin \cite{king2012ppcoin} to solve the energy waste and centralization issues. Coin age is defined as the number of a coin multiplied by the number of days it has been held by the holder. This incentivizes holders to hold coins for a longer time to earn more rewards. However, the mechanism can lead to coins be concentrated towards those who already hold large amounts coins, as they consistently earn more rewards. Similar to PoW, PoS also tends to become centralized over time \cite{he2020staking, tang2023pool}. Subsequently, other mechanisms have been proposed to improve blockchain technology, including Delegated Proof of Stake (DPoS), Practical Byzantine Fault Tolerance (PBFT) \cite{castro1999practical}, Proof of Activity (PoA) \cite{bentov2014proof}, etc. The improvements include reducing energy wastage, accelerating transaction processing speed, and decreasing centralization. However, none of them are sufficient for decentralization. For instance, some public blockchains including Solana\cite{yakovenko2018solana}, Aptos enhance processing speed by introducing centralized mechanisms. 
Solana has experienced several instances of downtime due to its centralized architecture. Similarly, the Aptos blockchain has only 120 validator nodes. With such a small number of nodes, it is not sufficiently decentralized. Therefore, these mechanisms cannot fully guarantee decentralization in practical scenarios.

Some distributed voting-based consensus algorithms including Voting-based Decentralized Consensus (VDC) \cite{sun2020voting}, RDV \cite{solat2018rdv} and Proof of Vote (PoV) \cite{li2017proof} have been proposed to improve decentralization while reducing energy waste. VDC \cite{sun2020voting} is a consensus algorithm that combines verifiable random functions (VRF). It employs a similar sub-user mechanism to Algorand, where VRF is used for selecting sub-users. By partitioning identities, VDC reduces the overhead of voting in large-scale networks, making it only suitable for more centralized consortium chains. RDV \cite{solat2018rdv} proposes a consensus algorithm that requires nodes to deposit coins before participating in consensus to prevent potential malicious behavior. Such a consensus mechanism still requires a BFT architecture to ensure that the majority of nodes are honest, as it does not prevent the potential for collusion among nodes to engage in dishonest voting behavior. PoV \cite{li2017proof} designs a distributed voting mechanism to ensure decentralization in blockchains. However, these voting-based consensus algorithms are challenging to apply to large-scale blockchain networks and are more suitable for consortium chains.

There are also some mechanisms based on verifiable functions have been proposed in \cite{raikwar2021r3v, gilad2017algorand}. R3V \cite{raikwar2021r3v} is a round-robin VDF-based consensus algorithm. It ensures decentralization by requiring stakeholders to solve VDF-based puzzles. However, there is no experiment analysis to prove its actual efficiency. Algorand \cite{gilad2017algorand} is a PoS consensus algorithm based on VRF, but it also faces similar centralization issues as other PoS algorithm. These mechanisms have not fully utilized the unpredictability of verifiable functions. The unpredictability of verifiable functions can be used to implement sufficient decentralized consensus algorithm. Therefore, this paper proposes a consensus mechanism that combines two verifiable functions and fully utilizes unpredictability.

The comparison of  the above consensus mechanisms based on shared characteristics is shown in Table \ref{related-works-comparison}. Energy consumption represents the amount of energy used by the consensus mechanism. The Matthew effect \cite{rigney2010matthew} describes whether the consensus mechanism leads to the centralization of tokens. The randomness is evaluated based on whether the mechanism incorporates random factors. Finally, centralization indicates whether the consensus mechanism is suitable for large-scale network.

\section{Preliminaries}
\label{preliminaries}

\subsection{Verifiable Random Function}
\label{preliminaries:VRF}

Verifiable Random Function
 (VRF) \cite{micali1999verifiable} is a public-key pseudorandom function that provides proofs that its outputs were correct. Only the owner of private key can produce the correct output and a corresponding proof using a message as input. Any entity can use the proof and the associated public key to check whether the output is valid. It consists of the following probabilistic polynomial-time algorithms.

\begin{itemize}
  \item $VRFKeyGen(1^\lambda) \rightarrow (pk, sk)$: On input a security parameter $1^\lambda$, it output a public-private key pair $(pk, sk)$.
  \item $VRFEval(sk, m) \rightarrow (r, \pi)$: On input the secret key $sk$ and the message $m$, it output a pseudo-random number $r$ and the proof $\pi$.
  \item $VRFVerify(pk, m, r, \pi) \rightarrow \{0,1\}$: It output 1 iff $r$ is the correct pseudo-random number produced by the evaluation algorithm on secret key $sk$ and message $m$.
\end{itemize}

\subsection{Verifiable Delay Function}
\label{preliminaries:VDF}

Verifiable Delay Function (VDF) \cite{boneh2018verifiable} is a time-lock puzzle that requires specified number of sequential steps to calculate. The VDF will produce a unique output that can be validated effectively and publicly based on the input. Therefore, the computation of VDF can only be performed serially, precluding the acceleration of computations through parallel computation. Dan Boneh introduced a simple VDF implementation \cite{boneh2018survey}, which produce the output by computing the formula as follows.

\begin{equation}
y = x^{2^T}
\label{vdf-calculate-equation}
\end{equation}

$y$ is the output produced based on the input $x$ under the time parameter $T$. The computation process requires $T$ rounds of sequential steps, with a time complexity of $O(T)$. While computing the VDF, the proof $\pi$ is simultaneously calculated. Any entity can check whether the output is valid with a time complexity of $O(log\ T)$ based on $\pi$ . The detail of the computation method is referenced in \cite{boneh2018survey}. It consists of the following probabilistic polynomial-time algorithms.

\begin{itemize}
  \item $VDFSetup(1^\lambda, T) \rightarrow pp$: On input a security parameter $1^\lambda$ and a time bound $T$, it output public parameters $pp$.
  \item $VDFEval(pp, x) \rightarrow (y, \pi)$: The primary VDF computation process. On input the public parameters $pp$ and an input $x$, it output the evaluation result $y$ and proof $\pi$.
  \item $VDFVerify(pp, x, y, \pi) \rightarrow \{0,1\}$: It outputs 1 iff $y$ is correct evaluation of the VDF on input $x$.
\end{itemize}

\subsection{Distributed clock synchronization}

Distributed clock synchronization algorithm can be implemented by executing Network Time Protocol (NTP) \cite{mills1985network} simultaneously between randomly composed pairs of entities. NTP is a client-server protocol which is widely used to implement clock synchronization in the Internet. The client send a synchronization request to a server with reliable clock, and adjust clock based on the response from the server. 

\begin{figure}[htbp]
  \centerline{\includegraphics[width=\linewidth]{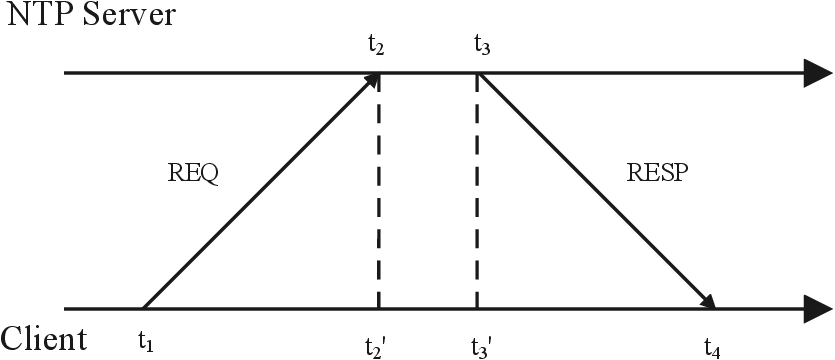}}
  \caption{NTP request-response process}
  \label{time-synchronization-fig}
\end{figure}

As shown in Figure \ref{time-synchronization-fig}, four time points can be obtained from request-response process in NTP. According to NTP, the clock offset $\theta$ between client and server can be calculated with the four time points as follows:

\begin{equation}
  \theta = \frac{(t_2-t_1) + (t_3-t_4)}{2}
  \label{NTP-theta-equations}
\end{equation}

The client can adjust its local clock according to the offset $\theta$, ultimately synchronizing the clock with the server. However, it is impossible to know which node is reliable in a distributed network. Nodes in a distributed network need to maintain a logical clock, and each node's logical clock within a small error margin of the global clock. The definitions of physical clock, logical clock and global clock are as follows.

\begin{itemize}
  \item \textbf{physical clock}: The system clock of a node is maintained by operating system. Applications can access the clock through system calls.
  \item \textbf{logical clock}: The clock obtained after eliminating clock offset. It is maintained by distributed applications, which combines the clock offset with the physical clock.
  \item \textbf{global clock}: A virtual clock representing the clock of the entire distributed network.
\end{itemize}

In distributed network, each node needs to maintain a real-time offset $\theta$, which represents the offset between the node's physical clock and the global clock. The real-time offset $\theta$ is obtained by each node periodically process NTP with randomly chosen neighboring node. The logical clock of a node will converge to the acceptable range of the global clock after a period of time. The experiments in Section \ref{evaluation} shown that the distributed clock synchronization algorithm base on NTP is reliable. 

\section{PoVF Consensus}
\label{povf-consensus}

PoVF is a consensus algorithm that dynamically selects a subset of nodes from the blockchain network as consensus nodes. Consensus nodes can collect transactions and propose blocks. It ensures that every node in the network has the opportunity to propose blocks, thereby making the PoVF-based blockchain sufficiently decentralized. To implement a fair consensus algorithm, consensus epoch is designed based on VDF and dynamically select consensus nodes within each epoch using VRF. 

\begin{figure*}[htbp]
  \centerline{\includegraphics[width=1\linewidth]{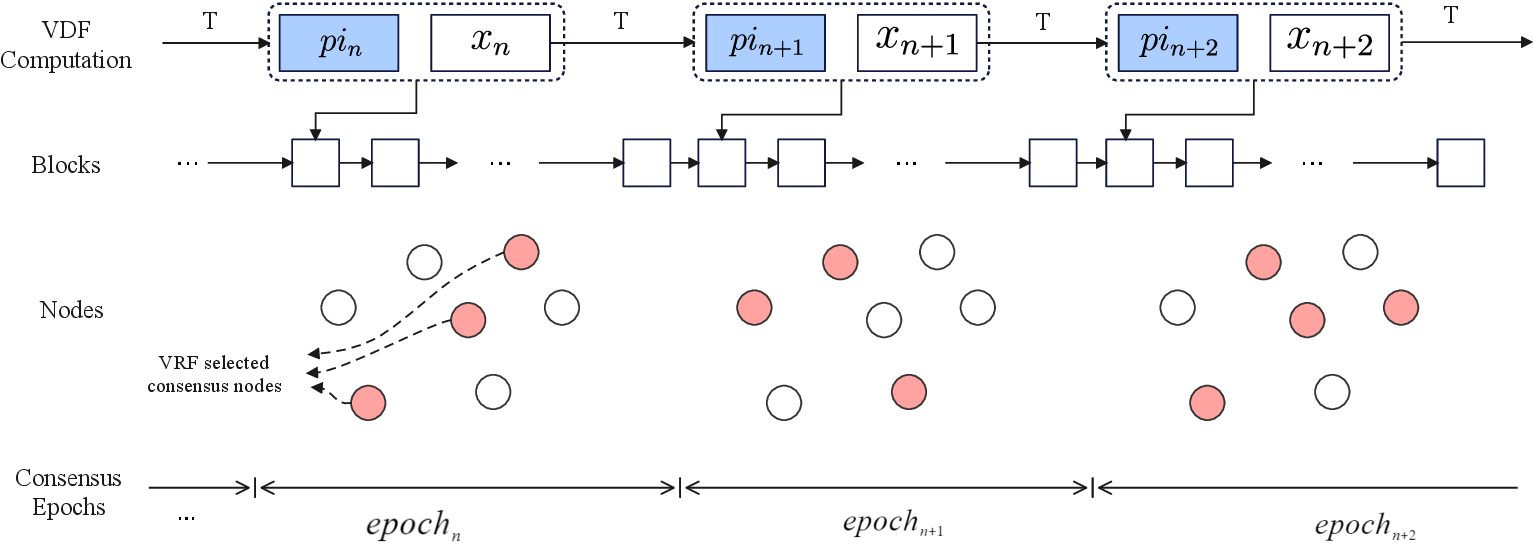}}
  \caption{PoVF consensus epoch overview}
  \label{consensus-epoch-fig}
\end{figure*}

The consensus epoch in PoVF is shown as Figure \ref{consensus-epoch-fig}. To achieve the functionality described above, PoVF has designed a special block structure shown in Section \ref{consensus-block-struct}. The consensus epoch for selecting a subset of nodes is described in Section \ref{consensus-consensus-epoch}. The valid nodes can be selected as consensus node by the node selection algorithm described in Section \ref{struct-node-selection}. To prevent potential Sybil attack in consensus epoch, the heartbeat mechanism is designed in Section \ref{consensus-register-update}. In addition, to prevent too many node be selected, the dynamic probabilistic mechanism is shown in Section \ref{struct-dynamic-prob}. Finally, the Delay buffer described in Section \ref{struct-delay-buffer} is used to ensure transactions are executed correct and avoid blockchain forks. The blocks broadcasted to the blockchain network will be confirmed after a certain delay. 

\subsection{Block Structure}
\label{consensus-block-struct}

\begin{table}[htbp]
  \caption{Block header attributes in PoVF}
  \begin{center}
  \begin{tabular}{m{2.5cm}m{4.5cm}}
  \hline
  \rule{0pt}{8pt}
  Attributes&  Description \\
  \hline
  \rule{0pt}{8pt}
  $timestamp$& Timestamp of when the block is proposed\\
  \hline
  \rule{0pt}{8pt}
  $prevBlockHash$& The block hash value of previous block\\
  \hline
  \rule{0pt}{8pt}
  $blockHash$& The block hash value of current block\\
  \hline
  \rule{0pt}{8pt}
  $merkleRoot$& Merkle root hash of the transactions included in the block\\
  \hline
  \rule{0pt}{8pt}
  $height$& The block height of current block\\
  \hline
  \rule{0pt}{8pt}
  $publicKey$& The public key of the block proposer\\
  \hline
  \rule{0pt}{8pt}
  $params$& The data related to VDF and VRF in PoVF\\
  \hline
  \end{tabular}
  \label{block-header-attributes}
  \end{center}
\end{table}

The block header attributes in PoVF are shown as Table \ref{block-header-attributes}. In addition to the regular attributes in block header, PoVF require a byte array $paramas$ to store the public parameters related to VDF and VRF. The $params$ in the genesis block and regular blocks correspond to the parameters in Table \ref{genesis-attributes-table} and Table \ref{regular-attributes-table} respectively.

In the genesis block, $params$ contains the public parameters of VDF, denoted as $pp = (N, l) \leftarrow VDFSetup(1^\lambda,T)$. In addition, the genesis node also need to set the expected maximum number of consensus nodes $\Omega$. Therefore, the genesis parameters are shown as Table \ref{genesis-attributes-table}, which are the public parameters that the genesis node must fill and include in the genesis block.

\begin{table}[htbp]
  \caption{Genesis parameters in PoVF}
  \begin{center}
  \begin{tabular}{cm{5cm}}
  \hline
  \rule{0pt}{8pt}
  Parameters&  Description \\
  \hline
  \rule{0pt}{8pt}
  $N$& The order of the integer group for $VDFEval$\\
  \hline
  \rule{0pt}{8pt}
  $l$& The auxiliary parameter used for calculating the proof during the $VDFEval$ computation\\
  \hline
  \rule{0pt}{8pt}
  $T$& The time parameter, specifying the number of rounds for $VDFEval$\\
  \hline
  \rule{0pt}{8pt}
  $t$& The time parameter, specifying the number of rounds for $VDFEval$ in heartbeat mechanism \\
  \hline
  \rule{0pt}{8pt}
  $t_{max}$ & The maximum response latency time in heartbeat mechanism \\
  \hline
  \rule{0pt}{8pt}
  $x_0$& 
  A pseudorandom number, used as the initial input for $VDFEval$\\
  \hline
  \rule{0pt}{8pt}
  $\Omega $& The expected maximum number of consensus nodes in PoVF\\
  \hline
  \end{tabular}
  \label{genesis-attributes-table}
  \end{center}
\end{table}

All blocks after the genesis block are regular blocks, which include parameters related to VRF. These VRF-related parameters are used to validate the block. Only nodes that produce a valid VRF output using the VDF output of the current epoch as input are eligible to propose blocks. The regular parameters included in the block are shown as Table \ref{regular-attributes-table}.

\begin{table}[htbp]
  \caption{Regular parameters in PoVF}
  \begin{center}
  \begin{tabular}{cm{5cm}}
  \hline
  \rule{0pt}{8pt}
  Parameters&  Description \\
  \hline
  \rule{0pt}{8pt}
  $x_i$& The VDF output in current i-th consensus epoch\\
  \hline
  \rule{0pt}{8pt}
  $pi_i$& The VDF proof related to $x_i$\\
  \hline
  \rule{0pt}{8pt}
  $r$& 
  The VRF pseudo-random number produced by the block proposer\\
  \hline
  \rule{0pt}{8pt}
  $\pi$& The VRF proof related to the random number $r$\\
  \hline
  \end{tabular}
  \label{regular-attributes-table}
  \end{center}
\end{table}

\subsection{Consensus Epoch}
\label{consensus-consensus-epoch}

As shown in Figure \ref{consensus-epoch-fig}, different subset of nodes are selected as consensus nodes in each consensus epoch based on the VDF output. The consensus nodes propose new block and broadcast to blockchain network. Meanwhile, consensus nodes can also update the $VDFEval$ output and include it in the new block. VDF requires an input $x_i$ and produce an output $x_{i+1}$ along with a proof $pi_{i+1}$ after computing for $T$ rounds in $i$-th epoch. The next epoch can be obtained by using the output $x_{i+1}$ of the VDF as the new input for the next round of computation. Each epoch produces an output that cannot be predicted in advance, which can be used to fairly select nodes. 

\begin{figure}[htbp]
  \renewcommand{\algorithmicrequire}{\textbf{Input:}}
  \renewcommand{\algorithmicensure}{\textbf{Output:}}
  \removelatexerror
  \begin{algorithm}[H]
      \caption{$VDFEval$ computation process}
      \KwIn {Initial input $x_0$, time parameter $T$, verify parameter $l$, order $N$}

      $x \leftarrow x_0$\;
      $i \leftarrow 0$\;
      \While{true}{
        $i \leftarrow i + 1$\; 
        $round \leftarrow 0$\;
        $pi \leftarrow 1, r \leftarrow 1$\;
        \For{$round < T$}{
          $b \leftarrow \lfloor 2r/l \rfloor$\;
          $r \leftarrow (2r\ mod \ l)$\;
          $pi \leftarrow (pi^2{x_{i - 1}}^b\ mod\ N)$\;
          $x \leftarrow (x^2\ mod\ N)$\;
          $round \leftarrow round + 1$\;
        }
        $x_i \leftarrow x$\;
        $pi_i \leftarrow pi$
      }
  \end{algorithm}
  \caption{VDF computation pseudo-code}
  \label{vdf-computation-fig}
\end{figure}

\begin{figure}[htbp]
  \renewcommand{\algorithmicrequire}{\textbf{Input:}}
  \renewcommand{\algorithmicensure}{\textbf{Output:}}
  \removelatexerror
  \begin{algorithm}[H]
      \caption{The $VDFVerify$ process}
      \KwIn {Previous input $x_{n-1}$, VDFEval output $x_n$, proof $pi_n$, time parameter $T$, verify parameter $l$, order $N$}
      \KwOut {Boolean result $correct$}

      $r \leftarrow (2^T\ mod\ l)$\;
      $h \leftarrow ({pi_n}^l{x_{n-1}}^r\ mod\ N)$\;
      $correct \leftarrow (x_n == h)$\;

      output $correct$\;
      
  \end{algorithm}
  \caption{VDF verification pseudo-code}
  \label{vdf-verification-fig}
\end{figure}

The pseudo-code for the $VDFEval$ computation as shown in Figure \ref{vdf-computation-fig}. Assuming the latest epoch is $epoch_n$. Consensus nodes need to obtain the latest result $(x_n, pi_n)$ and include it to the new block before packing transactions. As described in Section \ref{preliminaries:VDF}, VDF requires $T$ repetitions of computation, with a time complexity of $O(T)$. This ensures that no node can obtain the result in advance, guarantee that nodes cannot predict whether those nodes will be selected as consensus nodes. The result of $VDFEval$ computation can be verified quickly with time complexity of $O(log\ T)$ as the pseudo-code shown in Figure \ref{vdf-verification-fig}.

\begin{figure*}[htbp]
  \centerline{\includegraphics[width=0.95\linewidth]{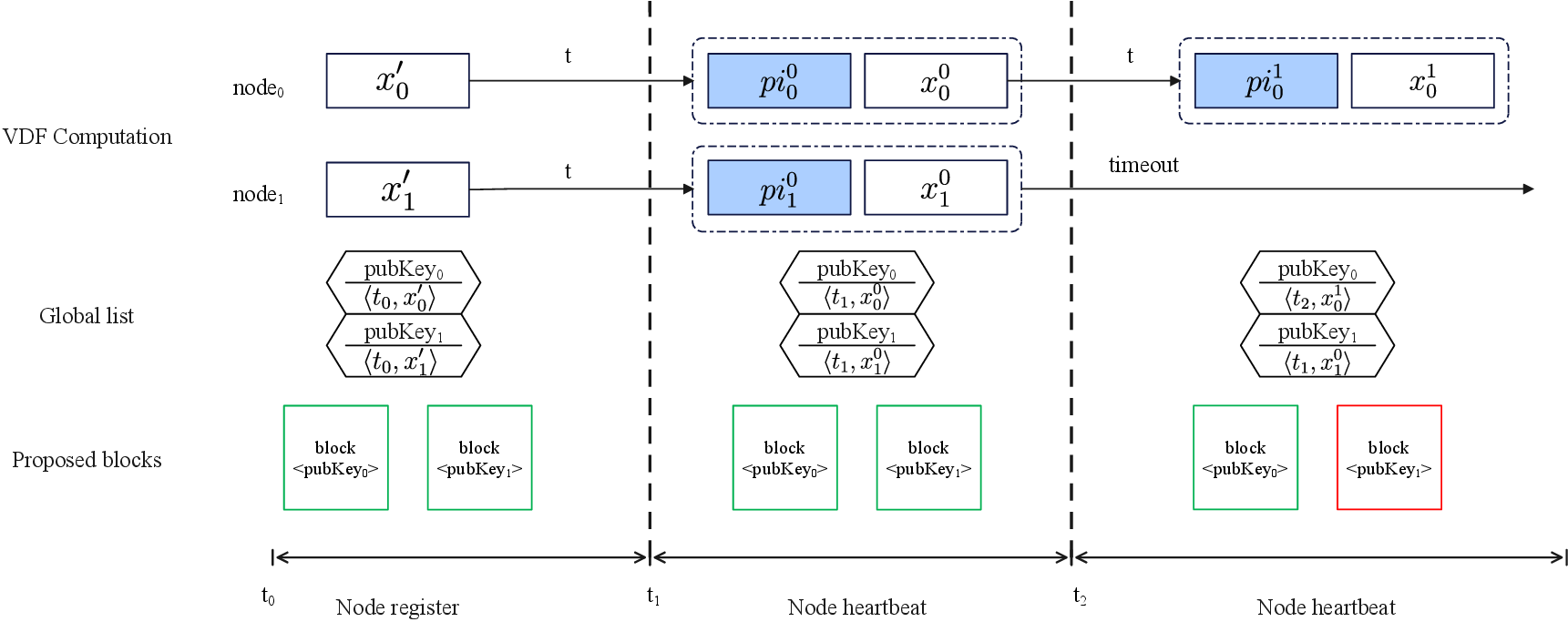}}
  \caption{PoW-like heartbeat mechanism overview}
  \label{vdf-heatbeat-fig}
\end{figure*}

\begin{example}
\label{example-vdf-epoch}
Select VDF computation parameters $(x_0, T, l, N) = (2, 10, 101, 41)$ as an example for ease of understanding. After one $VDFEval$ computation, the output is $x_1 = 5$ and $pi_1 = 5$. In the $VDFEval$ computation, $x_1 = 2^{2^{10}}\ mod\ 101 = 5$ and $pi_1 = 2^q \ mod\ 101 = 5$ where $q=\lfloor 2^{10}/41 \rfloor=24$. $(x_1, T, l, N)$ can be used as the input to start new $VDFEval$ round, which corresponds to a consensus epoch.

Using $(x_0, x_1, T, l, N) = (2, 5, 10, 101, 41)$ as input, the output of $VDFEval$ can be verified by $VDFVerify$. In $VDFVerify$ process, $r = 2^{10}\ mod\ 41 = 40$ and $h=5^{41}\cdot 2^{40}\ mod\ 101 = 5$.

In the above computation process, $2^T = ql + r$. While $VDFEval$ computes $pi = 2^q\ mod\ n$ during the calculation of $2^T\ mod\ n$, this enables the computation result to be quickly verified.
\end{example}

\subsection{Node Selection}
\label{struct-node-selection}

As Section \ref{consensus-consensus-epoch} described, each computation interval of the VDF constitutes a consensus epoch. And the consensus nodes are selected by the VDF output from the previous epoch. The node selection algorithm is designed to select a subset of nodes as consensus nodes in PoVF. VRF can generate a verifiable and uniformly distributed random number, which is crucial for the node selection algorithm. And the output of the VDF can be used as the input to the VRF, thereby generating an unpredictable random number in each consensus epoch. 

\begin{figure}[htbp]
  \renewcommand{\algorithmicrequire}{\textbf{Input:}}
  \renewcommand{\algorithmicensure}{\textbf{Output:}}
  \removelatexerror
  \begin{algorithm}[H]
      \caption{Node selection algorithm}
      \KwIn {VDF output $x_n$, consensus probability $p'$, private key $sk$}
      \KwOut {Boolean result $isConsensus$}

      $(r, \pi) \leftarrow VRFEval(sk, x_n)$\;
      $p \leftarrow \frac{r}{2^{randlen}}$\;

      $isConsensus \leftarrow (p \le p')$\;

      output $isConsensus$\;
      
  \end{algorithm}
  \caption{Node selection algorithm pseudo-code}
  \label{node-selection-fig}
\end{figure}

As shown in the pseudo-code in Figure \ref{node-selection-fig}, the n-th epoch VDF output $x_n$ be used as the input to VRF. Assuming that the VRF result computed by a node $i$ with its private key $sk_i$ is $(r, \pi) \leftarrow VRFEval(sk_i, x_n)$. Then the position of $r$ within its range can be computed as $p_i \leftarrow \frac{r}{2^{randlen}}$. As the mechanism described in Section \ref{struct-dynamic-prob}, the consensus probability $p'$ is dynamically set to adjust the proportion of consensus nodes in the entire blockchain network. If $p_i \le p'$, the the node $i$ is a consensus node in current epoch and can propose new block.

\begin{example}
  Assume there are two nodes in the network. In the 1-st consensus epoch, they run the node selection algorithm using $x_1$ from Example \ref{example-vdf-epoch} as input, with the outputs being $p_1 = 0.324$ and $p_2 = 0.986$, respectively. The consensus probability in the network is $p' = 0.9$, which means that only 10\% of the nodes are eligible to be selected as consensus nodes. Therefore, only the node with $p_2 = 0.986 > p'$ is selected as a consensus node in this example, and it can propose blocks to participate in the consensus.
\end{example}

\subsection{VDF heartbeat mechanism}
\label{consensus-register-update}

The node selection algorithm in PoVF is an offline validation algorithm, which cannot directly determine whether nodes have continuously participated in network maintenance.Therefore, potential adversaries could attempt to participate in proposals by generating a large number of identities. To defend against potential Sybil attacks, a heartbeat mechanism based on VDF is proposed in PoVF. The mechanism requires each node to maintain a global list $list$, which stores the timestamp of the most recent update from other nodes in the form of $\langle pk, x', st, timestamp, x\rangle$.

\begin{itemize}
  \item \textbf{Node register}: A new node calculate the random number $x' \leftarrow h(pk||st)$ with current logical clock $st$ by a hash function $h$. Then broadcast the tuple $\langle pk, x', st\rangle$ to blockchain network. Existing nodes in the blockchain network store this tuple to the $list$ upon receiving it. 
  \item \textbf{VDF calculate}: After joining the blockchain network, the new node starts computing $(x^0, pi^0) \leftarrow VDFEval(pp, x')$. Afterward, each computation requires the output of the previous round as input, starting with $(x^i, pi^i) \leftarrow VDFEval(pp, x^{i - 1})$.
  \item \textbf{Node heartbeat}: After each $VDFEval$ computation, nodes need to broadcast the tuple $\langle pk, x^i, pi^i\rangle$. Upon receiving this tuple, other nodes perform VDF verification, and update the corresponding $timestamp$ and $x^i$ in $list$ if the VDF result $(x^i, pi^i)$ is correct.
\end{itemize}

\begin{example}
  As shown in Figure \ref{vdf-heatbeat-fig}, global list $list$ is maintained by all node in blockchain network. $node_0$ and $node_1$ are register with tuples $\langle pk_0, x_0'\rangle$ and $\langle pk_1, x_1' \rangle$, and these tuples are stored in $list$ at the time point $t_0$. The blocks proposed by $node_0$ and $node_1$ respectively are both valid. At time point $t_1$, $node_0$ and $node_1$ both complete $VDFEval$ computation, resulting in new output $x_0^0$ and $x_1^0$. The outputs are broadcasted to the blockchain network and updated in $list$. The blocks proposed by $node_0$ and $node_1$ are both valid at this point.

A computation timeout occurs at time point $t_2$. $node_0$ completes the $VDFEval$ computation as usual, and the block it proposes still valid. However, if $node_1$ fails to complete the computation within $t_{max}$ defined in Table \ref{genesis-attributes-table}, resulting in the corresponding information in $list$ not being updated. Therefore, the block proposed by $node_1$ is invalid and will not be accepted by other nodes. If $node_1$ join the consensus again, it must register with new random seed and start a new computation.
\end{example}

In summary, the heartbeat mechanism requires each node to complete updates within $t_{max}$. The blocks proposed by non-updated nodes are invalid. This PoW-like mechanism effectively defends against potential Sybil attacks. The details of the security analysis will be described in Section \ref{security-analysis}.

\subsection{Dynamic probability}
\label{struct-dynamic-prob}

The numbers of nodes in the blockchain network gradually increase over time. Fixed probability for node selection would cause the number of consensus nodes to grow with the expansion of the network scale. Similar to the dynamic difficulty mechanism in PoW, PoVF requires a dynamic probability mechanism to ensure an upper limit on the number of consensus nodes.

As described in Section \ref{consensus-register-update}, each node maintains a global active nodes list $list$. Thus, the current number of valid nodes in the blockchain network can be obtained through $list$. Assuming that the number of valid nodes is $n$. And the maximum count of consensus nodes in the genesis block is $\Omega $. The adjustment formula for dynamic consensus probability can be defined as:

\begin{equation}
  prob(n) = min(\frac{\Omega}{n}, 1.0)
  \label{dynamic-probability}
\end{equation}

According to Equation \ref{dynamic-probability}, there are still consensus nodes even when there are few nodes in blockchain network. At the same time, it limit the number of consensus nodes when the scale is large. When $n \le \Omega$, all nodes in the blockchain network can propose blocks as consensus nodes. Only when $n > \Omega$, a part of the nodes will be selected as consensus nodes, and the number will be within a certain range of $\Omega$.

\begin{figure}[htbp]
  \centerline{\includegraphics[width=0.85\linewidth]{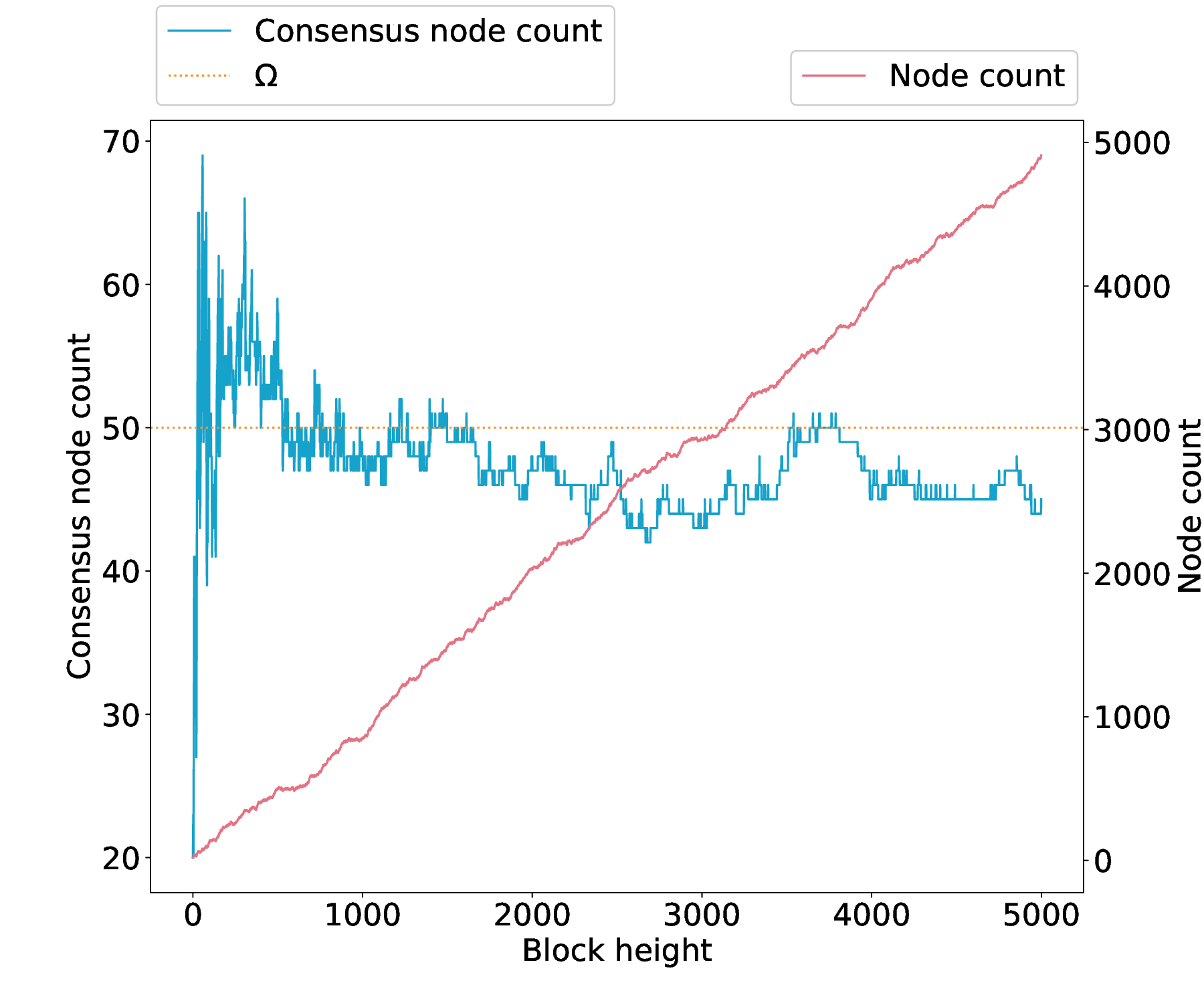}}
  \caption{Node selection simulate with dynamic probability}
  \label{consensus-simulate}
\end{figure}

The simulation experiment shown that the feasibility of dynamic selection probability. In the simulation experiment, $\Omega$ is fixed at 50, and nodes are continuously join or exit from blockchain network randomly. As shown in Figure \ref{consensus-simulate}, the number of consensus nodes does not increase with the expansion of the node scale, but fluctuates around $\Omega$.

\subsection{Delay buffer}
\label{struct-delay-buffer}

The delay in broadcasting blocks to the entire network increases with the expansion of the network scale. This often leads to short-live forks, where multiple different ledger copies exist in the network. Most blockchain systems solve the fork issue by block confirmation, such as Bitcoin requiring 6 blocks confirmation \cite{rosenfeld2014analysis} to ensure transactions are correct. Block confirmation can ensure that the probability of double-spending attacks occurring after a certain number of blocks is negligible. Additionally, block confirmation allows nodes in the entire network to have enough time to complete negotiations to ensure the consistency of the blockchain network \cite{biais2019blockchain}.

\begin{figure}[htbp]
  \centerline{\includegraphics[width=0.85\linewidth]{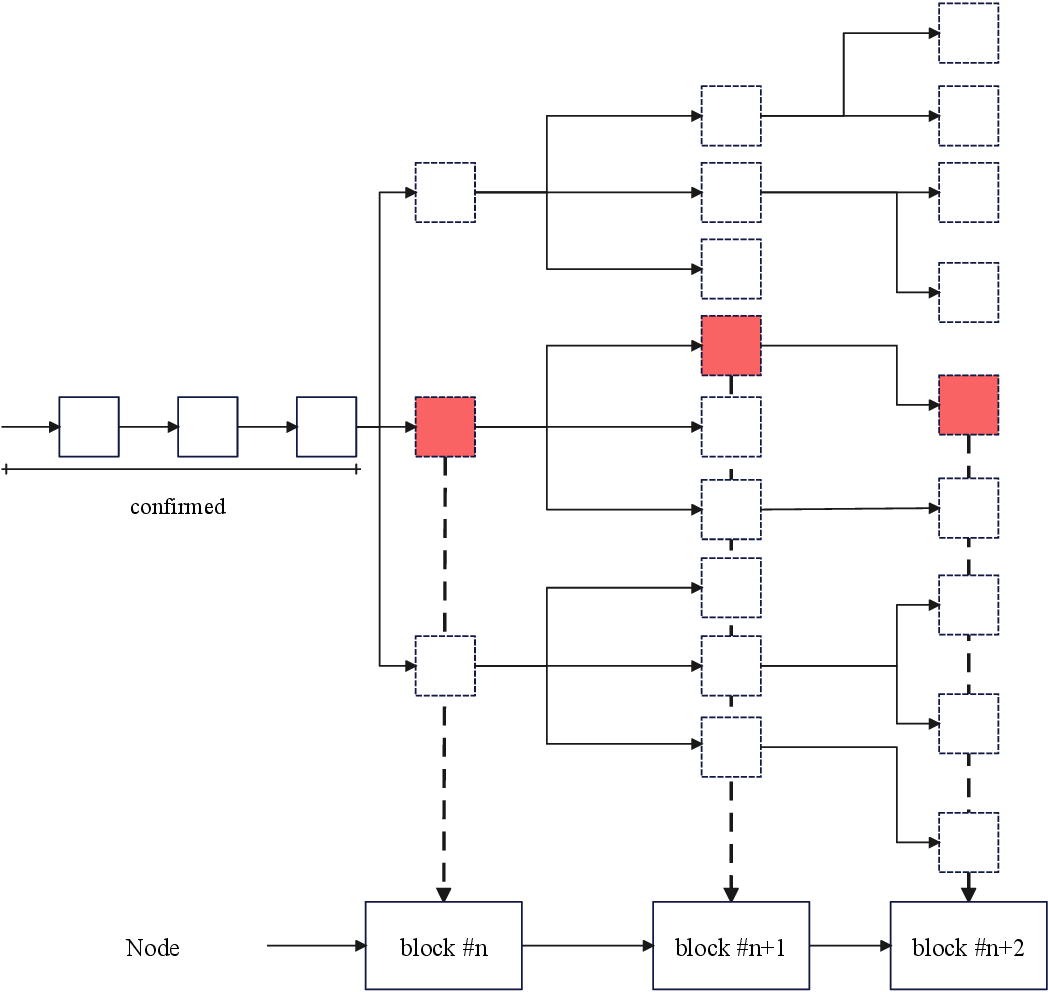}}
  \caption{Block buffer tree structure in blockchain node}
  \label{block-tree-view}
\end{figure}

To avoid potential the fork issue that can impact the communication and processing effectively of the blockchain network, the Delay buffer is proposed in this paper. The blocks received by each node actually form a tree structure based on parent-child relationships, as shown in Figure \ref{block-tree-view}. Each node eventually picks the block with the smallest timestamp and the largest number of transactions at each height. Therefore, given enough time to wait for blocks to arrive, it is possible to ensure that the block views of all nodes tend to converge. Nodes append received new blocks to the Delay buffer and select the best block to pop out after a certain height.

In this paper, the timestamp and transaction count of a block form a tuple $\langle timestamp, count\rangle$, which serves as the sorting criterion for blocks. Blocks with the same previous block are compared based on the tuple $\langle timestamp, count\rangle$. Blocks with smaller timestamps and more transactions have higher priority. To ensure timestamp consistency in the P2P network, the distributed clock synchronization algorithm mentioned in Section \ref{preliminaries} is used to synchronize the clocks between nodes. Each node continuously requests clock synchronization with neighboring nodes. Experimental results show that the clock deviation between nodes is within 1000ms. This makes that newly joined nodes can synchronize their clocks with the global clock of the entire network, ensuring the Delay buffer operation of the nodes.
 
\section{Security Analysis}
\label{security-analysis}

Public blockchains are open systems that allow  anyone to participate. It is necessary to consider the potential presence of malicious nodes in network, as they can compromise the security and decentralization of PoVF blockchain. The potential malicious behaviors are as follows:

\begin{itemize}
  \item \textbf{Sybil attack.} A malicious node generate a large number of public-private key pairs within current epoch, selecting valid pairs to propose numerous blocks. The attack exploits the vulnerability of offline verification in VRF, effectively increasing the probability of proposed blocks being selected.
  \item \textbf{Prophecy attack.} A malicious node has an advantage over other nodes in completing $VDFEval$ in advance. This allows it to predict whether nodes can be selected as consensus nodes. Based on this predictive ability, it can generate numerous valid public-private key pairs for the upcoming epoch and register in advance.
  \item \textbf{Replay attack.} A malicious node saves the outputs of the VDF in the heartbeat mechanism, and upon timeout or deliberately interrupting the computation, it re-registers with the same $x'$. This allows the malicious node to maintain the heartbeat without the need to compute the VDF and can also be used to forge multiple identities to implement a Sybil attack.
\end{itemize}

\subsection{Sybil attack}

Suppose an adversary $\mathcal{A}$ in a PoVF-based blockchain system want to propose numerous blocks to increase the probability of proposed blocks being selected. The adversary $\mathcal{A}$ needs to generate a large number of public-private key pairs within an epoch and select valid key pairs according to the rules described in Section \ref{struct-node-selection}. Moreover, $\mathcal{A}$ needs to quickly use these key pairs to propose blocks simultaneously. Otherwise, only blocks with smaller timestamps will be selected.

\begin{definition}
  \label{sybil-resistant-definition}
  (Sybil-resistant). We say the PoVF-based blockchain system is Sybil-resistant if the success probability of any polynomial-time adversary $\mathcal{A}$ is negligible in the following experiment:

  \begin{itemize}
    \item The number of adversary $\mathcal{A}$'s processor is denoted as $\nu_{\mathcal{A}}$, meaning it can simultaneously compute $\nu_{\mathcal{A}}$ tasks;
    \item The adversary $\mathcal{A}$ generates $n$ key pairs, corresponding to $n$ distinct node identities. Each identity takes time $t$ to complete the $VDFEval$ computation in one round of the heartbeat mechanism as described in Section \ref{consensus-register-update};
  \end{itemize}

  The adversary $\mathcal{A}$ is succeed if an only if, for any identity, the computation time is less than $t$ and satisfies $n > \nu_{\mathcal{A}}$.
\end{definition}

This property ensure that the chance for a polynomial-time adversary to increase its chance of participating in consensus by forging large identities is negligible. It also ensures that each node can maintain at most the number of identities equal to its parallel computing capability. In practical applications, the maximum number of identities a node can maintain simultaneously cannot exceed its CPU core count. We now show that PoVF satisfies the desired security requirement.

\begin{proof}
  The VDF algorithm used in PoVF is proposed by Wesolowski \cite{wesolowski2019efficient}. The algorithm was proven to be computed and finished in $O(\frac{t}{\nu_{\mathcal{A}}})$ with $\nu_{\mathcal{A}}$ processor. This means that the total time required to complete $n$ identities is $\frac{tn}{\nu_{\mathcal{A}}}$. To ensure that any identity completes the calculation in time $t$, it is required that $n \le \nu_{\mathcal{A}}$. Therefore, no adversary $\mathcal{A}$ can succeed in the experiment of Definition \ref{sybil-resistant-definition}.

  On the other hand, suppose an adversary $\mathcal{A}$ can success in experiment of Definition \ref{sybil-resistant-definition}, then it is possible to construct an adversary $\mathcal{A}'$ to finishing a $VDFEval$ computation less than $O(\frac{t}{\nu_{\mathcal{A}}})$. There is no such $\mathcal{A}'$ can finish a $VDFEval$ computation less than $O(\frac{t}{\nu_{\mathcal{A}}})$. Futhermore, such an adversary $\mathcal{A}$ can success in experiment of Definition \ref{sybil-resistant-definition} dose not exists. 
\end{proof}

In summary, the Sybil-resistant ensures that no adversary can forge a large number of identities to participate in consensus. This ensures PoVF is sufficiently fair and decentralization, with each node only able to maintain identities not exceeding its CPU core count, achieving real "one CPU one vote". It differs from PoW in that each identity is essentially independent, and the addition of a large number of identities does not guarantee adversaries a greater chance of participating in consensus.

\subsection{Prophecy attack}

Suppose an adversary $\mathcal{A}$ in a PoVF-based blockchain system seeks to predict the output of $VDFEval$ in advance. Furthermore, $\mathcal{A}$ utilizes the output to select valid public-private key pairs, as described in Section \ref{struct-node-selection} for the next epoch. These public-private key pairs can be registered in advance and used to propose blocks in the next epoch.

\begin{definition}
  (Unpredictable). We say the node selection mechanism is unpredictable iff the success probability of any polynomial-time adversary $\mathcal{A}$ is negligible in the following experiment:

  \begin{itemize}
    \item Assuming a consensus probability $p'$;
    \item A VDF oracle $\mathcal{O}$ initializes with a random message $m$ and security parameters $pp \leftarrow VDFSetup(1^\lambda, T)$. $\mathcal{O}$ has all VDF output $\mathcal{Y} = (y_0, y_1, ..., y_i, ...)$;
    \item $\mathcal{A}$ accesses $\mathcal{O}$ to obtain $y_\gamma  \in \mathcal{Y}$;
    \item $\mathcal{A}$ generates secret keys $\mathcal{K} = (sk_1, sk_2, ..., sk_n)$ and then submits to $\mathcal{O}$. Meanwhile, $n = poly(\frac{1}{p'})$ to ensure that $n$ is sufficiently large;
  \end{itemize}
  As described above, each secret key can be used to calculate a random number $r \leftarrow VRFEval(sk, m)$. Therefore, a secret key $sk_i \in \mathcal{K}$ corresponds to a random number $r_i \leftarrow VRFEval(sk_i, y_{\gamma + 1})$ with the next VDF output $y_{\gamma + 1}$. The random number set $\mathcal{R} = (r_1, r_2, ...,r_n)$ can be obtained by sequentially running $VRFEval$ on the elements in $\mathcal{K}$.Suppose that $m$ random number in $\mathcal{R}$ satisfies the conditions $\frac{r_i}{2^{randlen}} \le p'$. The adversary $\mathcal{A}$ is succeed if and only if $m$ satisfies the following condition:
  \begin{equation}
    \frac{m}{n} > p'
  \end{equation}
  \label{prophecy-attack-def}
\end{definition}

\begin{proof}
  Assuming an adversary $\mathcal{A}$ can succeed in Definition \ref{prophecy-attack-def}, then an adversary $\mathcal{A}'$ can be constructed through $\mathcal{A}$ to win in the game of distinguishing VDF output from a pseudo-random number.

  The adversary $\mathcal{A}'$ can access an oracle $\mathcal{O}$ in the game shown in Figure \ref{ind-vdf-game}. And the IND-VDF game shown in Figure \ref{ind-vdf-game} is described as follows:

  \begin{enumerate}
    \item The oracle $\mathcal{O}$ initial with a random message $x$ and security parameters $pp \leftarrow VDFSetup(1^{\lambda}, T)$ and $b \leftarrow \{0, 1\}$. $\mathcal{O}$ has all VDF output $\mathcal{Y}=(y_0,y_1,...,y_i,...)$. If $b=0$, each query to the oracle $\mathcal{O}$ returns $y_{\gamma} \in \mathcal{Y}$. Otherwise, the oracle $\mathcal{O}$ returns an random number $r \leftarrow G(1^\lambda)$ generated by the pseudo random function $G$;
    \item $\mathcal{A}$ access $\mathcal{A}'$, then $\mathcal{A}'$ access the oracle $\mathcal{O}$ to obtain $y_{\gamma} \in \mathcal{Y}$ and return to $\mathcal{A}$;
    \item $\mathcal{A}$ output secret key set $\mathcal{K} = (sk_1, sk_2, ..., sk_n)$ and query the constructed adversary $\mathcal{A}'$;
    \item $\mathcal{A}'$ receives the secret key set $\mathcal{K}$. If the conditions described in Definition \ref{prophecy-attack-def} are satisfies, then $\mathcal{A}'$ access the oracle $\mathcal{O}$ with $b'=0$. Otherwise, $\mathcal{A}'$ access the oracle with $b'=1$;
  \end{enumerate}

  \begin{figure}[htbp]
    \centerline{\includegraphics[width=\linewidth]{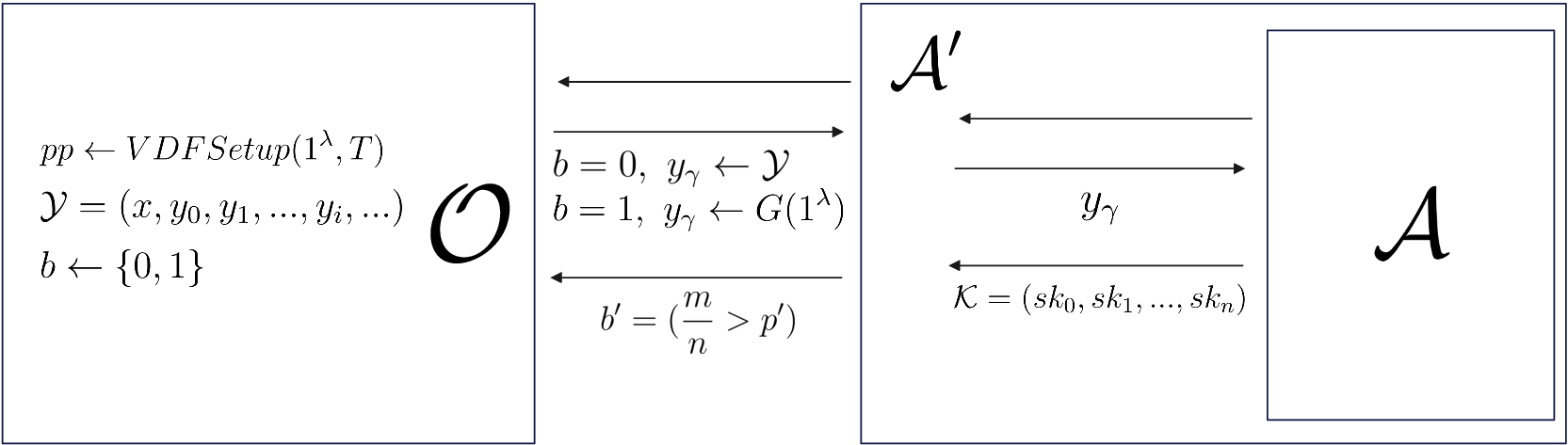}}
    \caption{IND-VDF game in PoVF}
    \label{ind-vdf-game}
  \end{figure}

  If such an adversary $\mathcal{A}$ described in Definition \ref{prophecy-attack-def} exists, then there exists a polynomial-time adversary $\mathcal{A}'$ in IND-VDF satisfies:

  \begin{equation}
    Pr[b=b'] > \frac{1}{2}
  \end{equation}

  According to the security proof of \cite{boneh2018verifiable}, no such adversary $\mathcal{A}'$ can succeed in IND-VDF due to the VDF being pseudo-random. Futhermore, such an adversary $\mathcal{A}$ that can succeed in Definition \ref{prophecy-attack-def} does not exists. 
\end{proof}

In summary, adversary can not predict the outcomes of $VDFEval$ computation in advance. The unpredictability of VDF ensures that all nodes cannot determine their roles for the next epoch until the $VDFEval$ computation is completed. This prevents potential malicious nodes from pre-selecting a sufficient number of valid key pairs to manipulate consensus.

\subsection{Replay attack}

Suppose an adversary $\mathcal{A}$ finished $n$ rounds heartbeat as described in Section \ref{consensus-register-update} and timeout at round $n + 1$. Assuming that the tuple in initial registration is $\langle pubKey_{\mathcal{A}}, x_{\mathcal{A}}, st \rangle$, the VDF output set is  $\mathcal{Y} = (y_0, y_1, ..., y_n)$. Since there was a timeout in the round $n+1$, the adversary $\mathcal{A}$ need to re-register with need random seed. However, it could potential engage in an attack by reusing the $x_0$ for registration and then replay the output set $\mathcal{Y}$ along with the corresponding proof without performing the $VDFEval$ computation. 

\begin{definition}
  \label{replay-resistant-def}
  (Replay-resistant). We say the PoVF-based blockchain system is replay-resistant if the success probability of any polynomial-time adversary $\mathcal{A}$ is negligible in the following experiment:

  \begin{itemize}
    \item $\mathcal{A}$ initializes with public-private key pair $(pk, sk) \leftarrow VRFKeyGen(1^{\lambda})$;
    \item $\mathcal{A}$ calculates $x_0 \leftarrow h(pk||t_0)$ with a selected logical clock $t_0$;
    \item $\mathcal{A}$ select a new logical clock $t_1 \neq t_0$, which is used to calculate $x_1 \leftarrow h(pk||t_1)$;
  \end{itemize}
  
  The adversary $\mathcal{A}$ succeeds if it managed to make $x_0$ equal to $x_1$. This property ensures that the probability for a malicious node successfully replaying $x'$ to register is negligible. Employing the timestamp as registration information ensures the unpredictability of the seed, thereby effectively preventing malicious nodes from registering with the same seed and Consequently mitigating malicious activities, thus avoiding malicious behavior.
\end{definition}

\begin{proof}
  The replay-resistant in PoVF is relays on the collision-resistant of the hash function $h$. In this paper, SHA-256 is used in PoVF to compute digests of various information. Therefor the replay-resistant in PoVF is relays on the collision-resistant of SHA-256.

  Suppose an adversary $\mathcal{A}$ can success in experiment of Definition \ref{replay-resistant-def}, then it is possible to construct a polynomial-time adversary $\mathcal{A}'$ to find another pre-image $m'$ of any message $m$  under SHA-256. There is no such $\mathcal{A}'$ that can success in probabilistic polynomial-time. Furthermore, such an adversary $\mathcal{A}$ in experiment of Definition \ref{replay-resistant-def} does not exists.
\end{proof}

In summary, no adversary can use the old VDF output in a new heartbeat round to achieve replay attack. Nodes in PoVF-based blockchain can only generate a new seed to restart the heartbeat after timeout. This ensures that all nodes can only use one identity to participate in the consensus. As a result, malicious node cannot accumulate enough identities to implement a Sybil attack.

\section{Performance evaluation}
\label{evaluation}

In this section, we evaluate the performance including throughput and clock synchronization offset in PoVF-based blockchain.

\subsection{Throughput}

\begin{figure*}[htbp]
  \centerline{\includegraphics[width=1\linewidth]{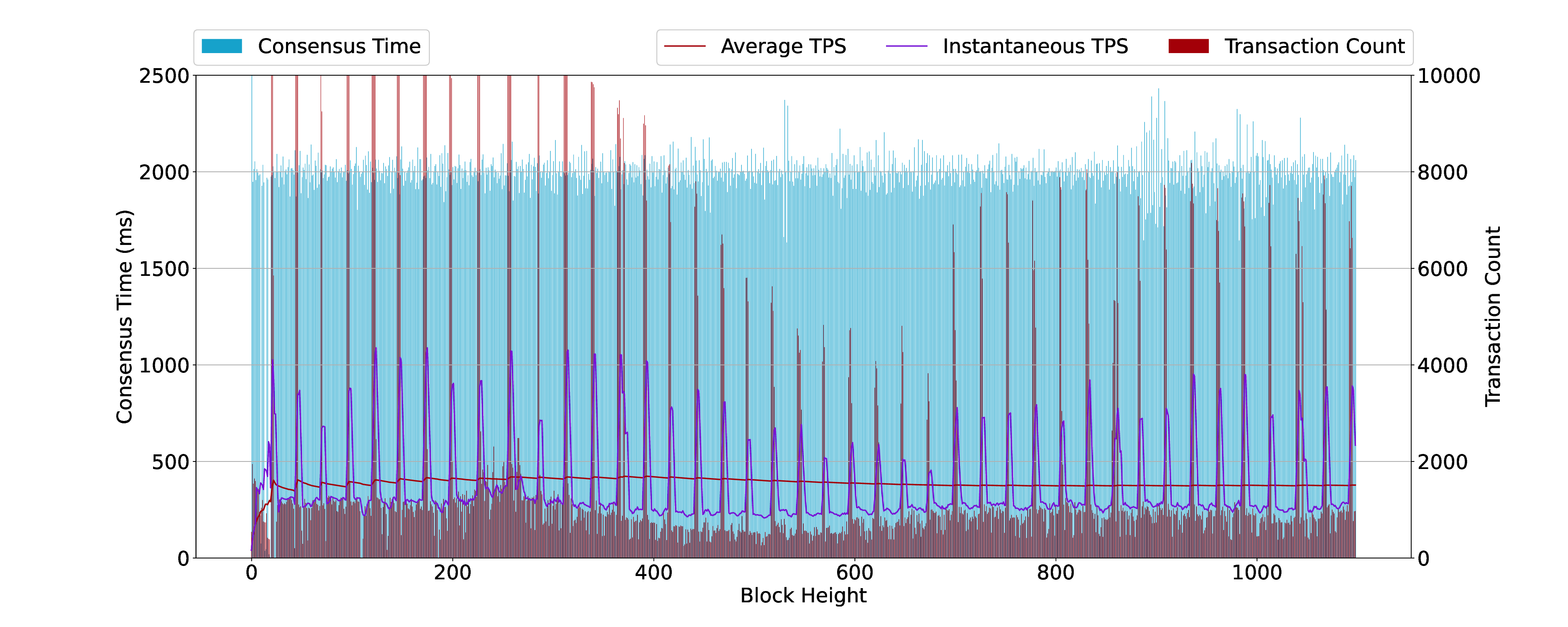}}
  \caption{Throughput result in PoVF-based blockchain}
  \label{norn-tps-figure}
\end{figure*}

A network with 200 nodes was set up for throughput evaluation. In the experiment, the height of the Delay buffer was set to 32 to ensure that all nodes had enough time to receive all blocks at the same height. Specifically, the other parameters for this test shown as in Table \ref{throughput-params-table}.

Over $3.3 \times 10^6$ transactions are constructed and sent to random node in blockchain network. Each node in the blockchain network, upon receiving and verifying the transactions, places them into its transaction pool and broadcasts them to other nodes. This ensures that all nodes in the blockchain network have enough transactions to execute. It simulates the scenario where the blockchain network receives a large number of transactions, facilitating a better assessment of the throughput of the PoVF-based blockchain. After approximately 1100 blocks, the throughput evaluation of the blockchain is shown in Figure \ref{norn-tps-figure}.

\begin{table}[htbp]
  \caption{Parameters for experiment}
  \begin{center}
  \begin{tabular}{m{2.5cm}m{0.8cm}m{3cm}}
  \hline
  \rule{0pt}{8pt}
  Parameters&  Value & Description \\
  \hline
  \rule{0pt}{8pt}
  Node count& 200 & The number of nodes in the blockchain network\\
  \hline
  \rule{0pt}{8pt}
  Delay height& 32 & The delay buffer size\\
  \hline
  \rule{0pt}{8pt}
  VDF round& $10^7$ & The number of $VDFEval$ computation rounds\\
  \hline
  \rule{0pt}{8pt}
  VRF probability& 10\% & The probability that a node is selected\\
  \hline
  \rule{0pt}{8pt}
  CPU core& 4 & The number of CPU core in a node\\
  \hline
  \rule{0pt}{8pt}
  Node memory& 4 GB& The memory in a node\\
  \hline
  \end{tabular}
  \label{throughput-params-table}
  \end{center}
\end{table}

The multi-axis chart in Figure \ref{norn-tps-figure} illustrates the time taken for block generation and the number of transactions at each height. The block generation time is calculated as the difference between the timestamps of consecutive blocks. Based on the information, the throughput of the blockchain network can be intuitively visualized. The parameter used to measure the throughput of the blockchain network is defined as TPS, representing the number of transactions the blockchain network can process per second. To better measure the TPS of a PoVF-based blockchain, the definitions of average TPS and instantaneous TPS are as follows.

\begin{figure*}[htbp]
  \centering
  \begin{minipage}[htbp]{0.495\textwidth}
    \subfloat[][CPU Usage]{\label{cpu-usage-figure}\includegraphics[width=1.1\textwidth]{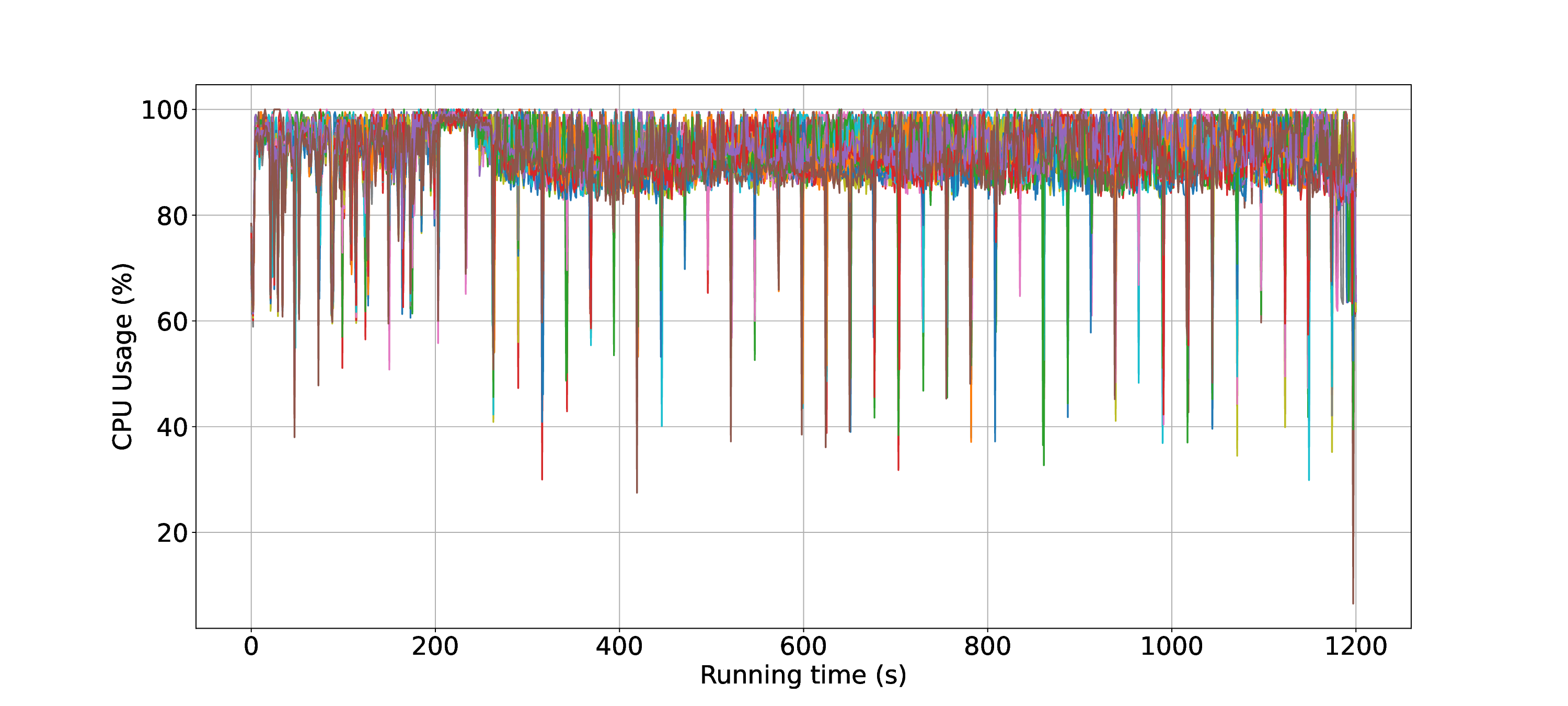}}
  \end{minipage}
  \begin{minipage}[htbp]{0.495\textwidth}
    \subfloat[][Memory Usage]{\label{mem-usage-figure}\includegraphics[width=1.1\textwidth]{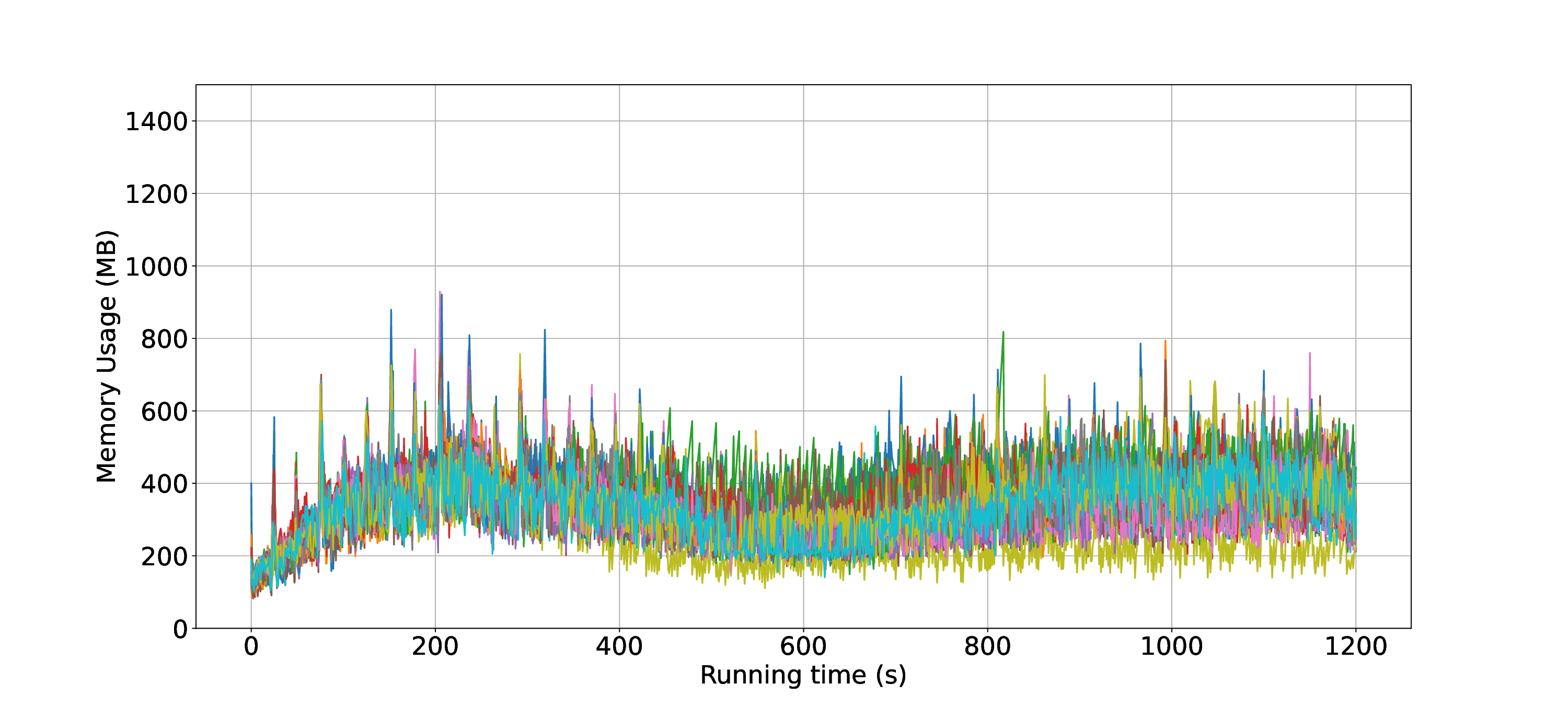}}
  \end{minipage}
  \caption{Resource consumption of the PoVF-based blockchain nodes}
  \label{resource-consumption-povf}
\end{figure*}

\begin{itemize}
  \item \textbf{Average TPS.} The average TPS is obtained by calculating the ratio of the total number of transactions processed up to a certain block to the elapsed time, which corresponds to the overall transaction processing capacity of the blockchain network.
  \item \textbf{Instantaneous TPS.} The instantaneous TPS is the ratio of the number of transactions in adjacent blocks to the time elapsed between them. It represents the transaction processing capacity of the blockchain network at a specific moment. With instantaneous TPS, the maximum processing capacity of the blockchain network can be effectively measured.
\end{itemize}

As shown in Figure \ref{norn-tps-figure}, the x-axis represents the block height. The left y-axis represents the consensus time for blocks at each height, corresponding to the cyan bar in the figure. Consensus time is calculated by measuring the time difference between the timestamp of a block at each height and that of the previous block. According to Figure \ref{norn-tps-figure}, the average consensus time for each block is 2 seconds. 

The right y-axis shows the number of transactions per block at each height, corresponding to the red bar. The average TPS and instantaneous TPS are calculated by combining consensus time and transaction count at each block height. The red line in Figure \ref{norn-tps-figure} represents the average TPS of the blockchain network in this experiment. The average TPS of the PoVF-based blockchain implemented in this paper is $1.5 \times 10^3$. In addition, the purple line represents the instantaneous TPS of the blockchain network in this experiment. The maximum instantaneous TPS reached $4.3 \times 10^3$, which is the maximum processing capacity of the PoVF-based blockchain network when receiving a large number of transactions. In the adjacent blocks where the instantaneous TPS reached its maximum, the number of transactions reached the block's transaction limit. This indicates that in this evaluation, the blockchain network has reached its maximum processing capacity limit. Through this throughput evaluation, the PoVF-based blockchain demonstrates high TPS and the capability to handle a large number of transactions.

The resource consumption of the PoVF-based blockchain is shown in Figure \ref{resource-consumption-povf}, which illustrates the CPU and memory overhead for all nodes in this experiment. According to Figure \ref{cpu-usage-figure}, the CPU usage of all nodes in this experiment is close to 100\%. Meanwhile, as shown in Figure \ref{mem-usage-figure}, the memory overhead for all nodes remained below 1GB. Based on the results of this experiment, the PoVF-based blockchain nodes have reached their processing limit under the current configuration. In the experimental environment with only 4 CPU cores, the average TPS remains around $1.5 \times 10^3$.

\begin{figure}[htbp]
  \centerline{\includegraphics[width=1\linewidth]{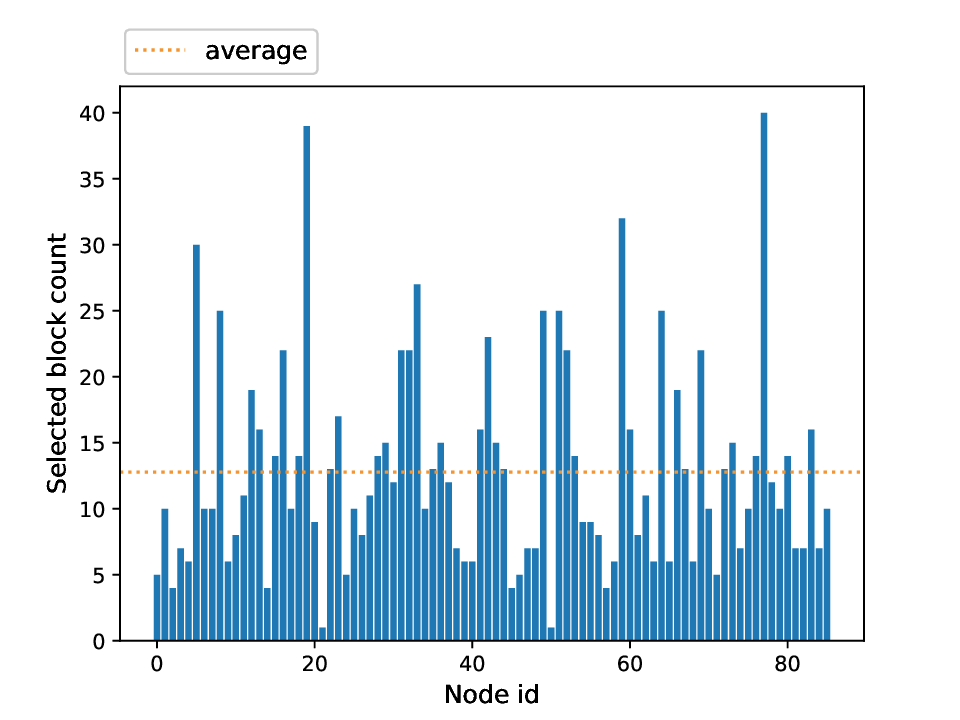}}
  \caption{Selected block distribution}
  \label{selected-distribution-figure}
\end{figure}

Additionally, data on the number of blocks proposed and selected by each node were collected. As shown in Figure \ref{selected-distribution-figure}, each bar represents the number of valid blocks proposed by nodes, with an average of 13 valid blocks proposed per node. The data shown in the figure indicates that all nodes in the PoVF-based blockchain have the opportunity to participate in consensus. This effectively demonstrates the decentralization of PoVF.

In summary, the evaluation of the PoVF-based blockchain demonstrates its robust transaction processing capability. Its peak TPS can reach $4.2 \times 10^3$, with an average TPS of $2 \times 10^3$ under typical conditions. Moreover, PoVF achieves high performance while ensuring decentralized and fair consensus. Based on the distribution of valid blocks proposed by each node, it is evident that all nodes have the opportunity to participate in consensus and propose blocks.

\subsection{Clock synchronization}

In PoVF, accurate timestamps are crucial for the nodes in the blockchain network. When proposing block, nodes need to include timestamp into block. In addition, timestamp is used to determine the validity of blocks and to decide the priority of blocks at the same height. The distributed clock synchronization algorithm is employed to synchronize the logical clocks between nodes, ensuring that these logical clocks tend towards a unified global clock.

\begin{figure}[htbp]
  \centerline{\includegraphics[width=1.1\linewidth]{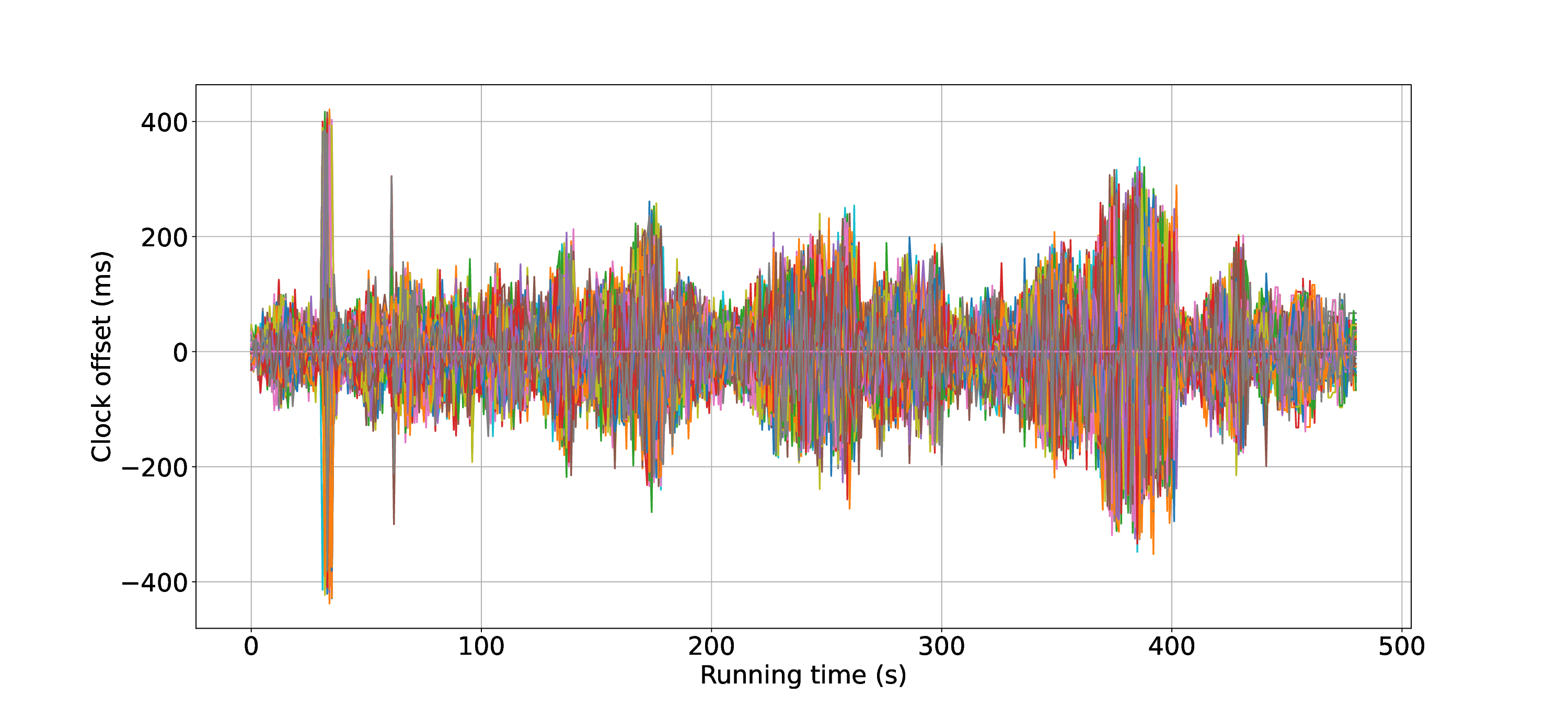}}
  \caption{Clock synchronization offset}
  \label{clock-sync-offset}
\end{figure}

The PoVF-based blockchain implemented in this paper has Prometheus\footnote{Prometheus is a open source systems and service monitoring system at https://github.com/prometheus/prometheus} metric monitoring functionality. Each time a node requests synchronization, the clock offset calculated is logged. Therefore, the offset in distributed time synchronization can be obtained through metrics. Clock synchronization offset are collected automatically by Prometheus while evaluating the throughput. As described above, the PoVF-based blockchain system generate approximately 700 blocks for evaluation. The time synchronization offset of all 200 blockchain nodes was collected. After processing the collected data, the clock synchronization offset of all nodes at each time point is shown in Figure \ref{clock-sync-offset}.

As shown in Figure \ref{clock-sync-offset}, in the blockchain network of 200 nodes, the maximum clock offset between nodes does not exceed 900ms. Such an offset is sufficient to ensure the blockchain network works correctly, since the average block generation time is 2s, which is much larger than 900 ms. This ensure the correct operation of nodes and the validity of the blocks they propose. Nodes in the blockchain network can also correctly select the block with the highest priority.
\subsection{Comparison}

PoVF is a sufficiently decentralized consensus mechanism. To compare with other consensus mechanisms, random consecutive blocks are sampled on different blockchain including Bitcoin, Ethereum and Solana, etc. Subsequently, we separately tallied the number of blocks proposed by each proposer across different blockchains. The number of blocks data defined as the following formula, $b_i$ is the number of blocks propose by proposer $i$.

\begin{equation}
  \mathcal{B} = \{b_1, b_2, ..., b_n\}
\end{equation}

In this experiment, the standard deviation (SD) is used to measure the dispersion in the number of blocks proposed by nodes across different blockchains. However, in the most extreme case where one node proposes all blocks, the overall SD would be very small. To better assess the decentralization level, the Gini coefficient (GC) of each blockchain is calculated to measure the decentralization of nodes participating in consensus.

\begin{itemize}
  \item \textbf{Standard deviation.} The SD can be used to measure the dispersion of a set of data. The SD across different blockchains can be calculated as the following formula.
  \begin{equation}
    \sigma = \sqrt{\frac{1}{n}\sum\limits_{i=1}^n (b_i - \overline{b})^2},where \ \overline{b} = \frac{1}{n}\sum\limits_{i=1}^nb_i
  \end{equation}
  \item \textbf{Gini coefficient.} The GC can be used to measure the decentralization of a set of data. It is typically used to assess the degree of income disparity among residents. The gini coefficient across different blockchains can be calculated as the following formula.
  \begin{equation}
    G = \frac{\sum\limits_{i=1}^{n}\sum\limits_{j=1}^n |b_i - b_j|}{2\sum\limits_{i=1}^{n}\sum\limits_{j=1}^nb_j}
  \end{equation}
\end{itemize}

\begin{figure}[htbp]
  \centerline{\includegraphics[width=0.85\linewidth]{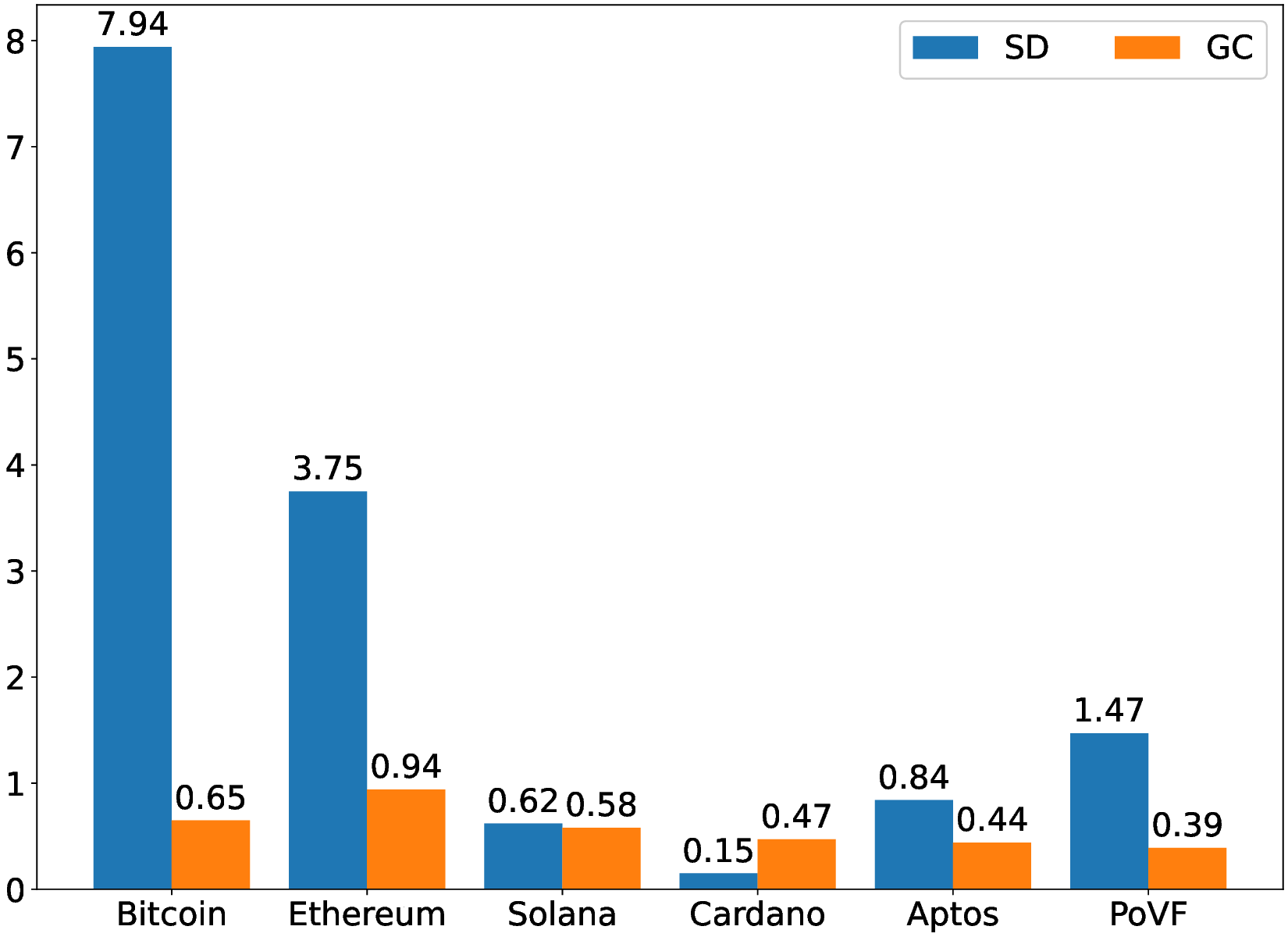}}
  \caption{Decentralization indicator of difference blockchains}
  \label{povf-comparison-figure}
\end{figure}

After completing the sampling of various blockchain systems, two metrics were calculated as shown in Figure \ref{povf-comparison-figure}. The blue bar represents the standard deviation. Bitcoin, Ethereum, and PoVF having standard deviations of 7.94, 3.75, and 1.47, respectively, which are relatively high compared to other blockchain systems. This is mainly due to the smaller impact of the number of nodes participating in the consensus. In Contrast, Solana, Cardano, Aptos and PoVF results in a much lower SD. It should be noted that the relatively small network size in PoVF experiments leads to a higher standard deviation.

Therefore, the Gini coefficient is used to further analyze the decentralization of consensus mechanisms. The Gini coefficient quantifies the recognition of blocks proposed by blockchain nodes and is independent of the number of nodes in the sampled data. As shown in Figure \ref{povf-comparison-figure}, the GC values for Ethereum, Bitcoin, and Solana are 0.94, 0.65, and 0.58, respectively, indicating a higher degree of centralization for these networks. In contrast, the GC for PoVF is only 0.39, the smallest among all the sampled blockchain systems. In summary, if the PoVF-based blockchain has a lager network, the probability of all nodes participating in consensus will be equal. Futhermore, the SD and GC obtained by sampling its blocks would be lower. The comparison shows that PoVF is the more decentralized consensus mechanism, which ensures the decentralization of all nodes participating in blockchain consensus.

\section{Conclusion}
\label{conclusion}

In this paper, we present PoVF, a decentralized consensus mechanism based on verifiable functions. PoVF combines VDF and VRF, provides a new consensus mechanism for selecting a subset of nodes in the blockchain network to propose blocks. VDF provides unpredictability and pseudo-randomness, which guarantee the selection result is unpredictable. In addition, VRF provides verifiability and pseudo-randomness, allowing nodes to discern the legitimacy of a new block without the need for excessive additional communication. PoVF effectively exploits the properties of verifiable functions within the consensus mechanism, thereby achieving a decentralized and efficient consensus mechanism.

This paper also considers potential problems with PoVF in practical implementations. For instance, fixed probability for node selection would cause the number of consensus nodes to grow with the expansion of the network scale. Therefore, a dynamic probability adjustment algorithm is proposed to limit the number of consensus nodes. And the broadcast latency of P2P networks that can lead to blockchain forks. To ensure blocks at the same height are all fully broadcast into P2P network, the Delay buffer is designed to delay block confirmation. It avoids blockchain forks caused by broadcast delays in the blockchain network.In addition, a VDF-based PoW-like heartbeat mechanism is used to defend against potential Sybil attacks. With security analysis, PoVF is provably secure. It can resistant to potential Sybil attacks, prophecy attack and replay attack. Indeed, since it's based on the P2P network, it can also withstand DDoS attacks. According to the experiment, PoVF provides sufficient decentralization while offering significant performance. The peak TPS can reach $4 \times 10^3$, surpassing the throughput of most existing public blockchains. It can be concluded that PoVF achieves a higher degree of decentralization and provides a fairer consensus mechanism through the comparison. 

The future research directions of this paper include a more detailed analysis of the Delay Buffer, fairness metrics for consensus mechanisms, and precise time synchronization algorithm. The Delay Buffer is influenced by various factors, such as network scale, block size, and block broadcast delay, etc. It is necessary to find the impact of these factors and calculate a more accurate delay height. In this paper, decentralization is analyzed by using standard deviation and Gini coefficient. However, fairness is influenced by multiple factors, such as the share of mining pools with higher hash rates, the share of capital, the likelihood of nodes participating in consensus, and the design of the incentive mechanism, etc. Combining various factors to obtain an objective metric can provide a thorough analysis of consensus mechanisms. We believe this is an important and challenging research direction. Additionally, implementing precise time synchronization algorithms to ensure consistency in operations executed by each node is challenging to further improving the real-time performance of the blockchain.

\section*{Acknowledgements}
\noindent
This research was supported by the National Natural Science Foundation of China, Grant No (U2033212) and the Sichuan Science and Technology Program under grant No. 2022JDRC0006.

\bibliographystyle{elsarticle-num} 
\bibliography{ref}

\begin{thebibliography}{10}
\expandafter\ifx\csname url\endcsname\relax
  \def\url#1{\texttt{#1}}\fi
\expandafter\ifx\csname urlprefix\endcsname\relax\def\urlprefix{URL }\fi
\expandafter\ifx\csname href\endcsname\relax
  \def\href#1#2{#2} \def\path#1{#1}\fi

\bibitem{nakamoto2008bitcoin}
S.~Nakamoto, Bitcoin: A peer-to-peer electronic cash system, Decentralized business review (2008).

\bibitem{feng2024pbag}
X.~Feng, K.~Cui, L.~Wang, Z.~Liu, J.~Ma, Pbag: A privacy-preserving blockchain-based authentication protocol with global-updated commitment in iovs, IEEE Transactions on Intelligent Transportation Systems (2024).

\bibitem{wang2020blockchain}
Q.~Wang, X.~Zhu, Y.~Ni, L.~Gu, H.~Zhu, Blockchain for the iot and industrial iot: A review, Internet of Things 10 (2020) 100081.

\bibitem{chohan2021double}
U.~W. Chohan, The double spending problem and cryptocurrencies, Available at SSRN 3090174 (2021).

\bibitem{huang2021rich}
Y.~Huang, J.~Tang, Q.~Cong, A.~Lim, J.~Xu, Do the rich get richer? fairness analysis for blockchain incentives, in: Proceedings of the 2021 international conference on management of data, 2021, pp. 790--803.

\bibitem{sai2021taxonomy}
A.~R. Sai, J.~Buckley, B.~Fitzgerald, A.~Le~Gear, Taxonomy of centralization in public blockchain systems: A systematic literature review, Information Processing \& Management 58~(4) (2021) 102584.

\bibitem{ren2019pooled}
L.~Ren, P.~A. Ward, Pooled mining is driving blockchains toward centralized systems, in: 2019 38th International Symposium on Reliable Distributed Systems Workshops (SRDSW), IEEE, 2019, pp. 43--48.

\bibitem{wheeb2020simulated}
A.~H. Wheeb, D.~N. Kanellopoulos, Simulated performance of sctp and tfrc over manets: The impact of traffic load and nodes mobility, International Journal of Business Data Communications and Networking (IJBDCN) 16~(2) (2020) 69--83.

\bibitem{king2012ppcoin}
S.~King, S.~Nadal, Ppcoin: Peer-to-peer crypto-currency with proof-of-stake, self-published paper, August 19~(1) (2012).

\bibitem{shifferaw2021limitations}
Y.~Shifferaw, S.~Lemma, Limitations of proof of stake algorithm in blockchain: A review, Zede Journal 39~(1) (2021) 81--95.

\bibitem{deirmentzoglou2019survey}
E.~Deirmentzoglou, G.~Papakyriakopoulos, C.~Patsakis, A survey on long-range attacks for proof of stake protocols, IEEE access 7 (2019) 28712--28725.

\bibitem{gavzi2018stake}
P.~Ga{\v{z}}i, A.~Kiayias, A.~Russell, Stake-bleeding attacks on proof-of-stake blockchains, in: 2018 Crypto Valley conference on Blockchain technology (CVCBT), IEEE, 2018, pp. 85--92.

\bibitem{li2020decentralized}
C.~Li, P.~Li, D.~Zhou, Z.~Yang, M.~Wu, G.~Yang, W.~Xu, F.~Long, A.~C.-C. Yao, A decentralized blockchain with high throughput and fast confirmation, in: 2020 $\{$USENIX$\}$ Annual Technical Conference ($\{$USENIX$\}$$\{$ATC$\}$ 20), 2020, pp. 515--528.

\bibitem{he2020staking}
P.~He, D.~Tang, J.~Wang, Staking pool centralization in proof-of-stake blockchain network, Available at SSRN 3609817 (2020).

\bibitem{tang2023pool}
D.~Tang, P.~He, Z.~Fan, Y.~Wang, Pool competition and centralization in pos blockchain network, Applied Economics (2023) 1--20.

\bibitem{castro1999practical}
M.~Castro, B.~Liskov, et~al., Practical byzantine fault tolerance, in: OsDI, Vol.~99, 1999, pp. 173--186.

\bibitem{bentov2014proof}
I.~Bentov, C.~Lee, A.~Mizrahi, M.~Rosenfeld, Proof of activity: Extending bitcoin's proof of work via proof of stake [extended abstract] y, ACM SIGMETRICS Performance Evaluation Review 42~(3) (2014) 34--37.

\bibitem{yakovenko2018solana}
A.~Yakovenko, Solana: A new architecture for a high performance blockchain v0. 8.13, Whitepaper (2018).

\bibitem{sun2020voting}
G.~Sun, M.~Dai, J.~Sun, H.~Yu, Voting-based decentralized consensus design for improving the efficiency and security of consortium blockchain, IEEE Internet of Things Journal 8~(8) (2020) 6257--6272.

\bibitem{solat2018rdv}
S.~Solat, Rdv: An alternative to proof-of-work and a real decentralized consensus for blockchain, in: Proceedings of the 1st Workshop on Blockchain-enabled Networked Sensor Systems, 2018, pp. 25--31.

\bibitem{li2017proof}
K.~Li, H.~Li, H.~Hou, K.~Li, Y.~Chen, Proof of vote: A high-performance consensus protocol based on vote mechanism \& consortium blockchain, in: 2017 IEEE 19th International Conference on High Performance Computing and Communications; IEEE 15th International Conference on Smart City; IEEE 3rd International Conference on Data Science and Systems (HPCC/SmartCity/DSS), IEEE, 2017, pp. 466--473.

\bibitem{raikwar2021r3v}
M.~Raikwar, D.~Gligoroski, R3v: Robust round robin vdf-based consensus, in: 2021 3rd Conference on Blockchain Research \& Applications for Innovative Networks and Services (BRAINS), IEEE, 2021, pp. 81--88.

\bibitem{gilad2017algorand}
Y.~Gilad, R.~Hemo, S.~Micali, G.~Vlachos, N.~Zeldovich, Algorand: Scaling byzantine agreements for cryptocurrencies, in: Proceedings of the 26th symposium on operating systems principles, 2017, pp. 51--68.

\bibitem{rigney2010matthew}
D.~Rigney, The Matthew effect: How advantage begets further advantage, Columbia University Press, 2010.

\bibitem{micali1999verifiable}
S.~Micali, M.~Rabin, S.~Vadhan, Verifiable random functions, in: 40th annual symposium on foundations of computer science (cat. No. 99CB37039), IEEE, 1999, pp. 120--130.

\bibitem{boneh2018verifiable}
D.~Boneh, J.~Bonneau, B.~B{\"u}nz, B.~Fisch, Verifiable delay functions, in: Annual international cryptology conference, Springer, 2018, pp. 757--788.

\bibitem{boneh2018survey}
D.~Boneh, B.~B{\"u}nz, B.~Fisch, A survey of two verifiable delay functions, Cryptology ePrint Archive (2018).

\bibitem{mills1985network}
D.~L. Mills, Network time protocol (ntp), Tech. rep. (1985).

\bibitem{rosenfeld2014analysis}
M.~Rosenfeld, Analysis of hashrate-based double spending, arXiv preprint arXiv:1402.2009 (2014).

\bibitem{biais2019blockchain}
B.~Biais, C.~Bisiere, M.~Bouvard, C.~Casamatta, The blockchain folk theorem, The Review of Financial Studies 32~(5) (2019) 1662--1715.

\bibitem{wesolowski2019efficient}
B.~Wesolowski, Efficient verifiable delay functions, in: Advances in Cryptology--EUROCRYPT 2019: 38th Annual International Conference on the Theory and Applications of Cryptographic Techniques, Darmstadt, Germany, May 19--23, 2019, Proceedings, Part III 38, Springer, 2019, pp. 379--407.

\end{thebibliography}
\end{document}